\documentclass[12pt]{iopart}
\usepackage{iopams}

\expandafter\let\csname equation*\endcsname\relax

\expandafter\let\csname endequation*\endcsname\relax

\usepackage{amsfonts}
\usepackage{amsmath}
\usepackage{amsthm}
\usepackage{mathrsfs}
\usepackage{amscd}
\usepackage{amssymb}
\usepackage{amsxtra}
\usepackage{bm}           
\usepackage{bbm}
\usepackage{graphicx}

\usepackage[colorlinks]{hyperref}

\usepackage[sort&compress,numbers,square]{natbib}

\newcommand{\la}{\langle}
\newcommand{\ra}{\rangle}
\newcommand{\bra}[1]{\ensuremath{\langle#1|}}
\newcommand{\ket}[1]{\ensuremath{\left|#1\right\rangle}}
\newcommand{\braket}[2]{\ensuremath{\langle #1|#2\rangle}}

\newcommand{\mean}[1]{\ensuremath{\left\langle #1 \right\rangle}}

\newcommand{\cL}{\mathcal{L}}
\newcommand{\cD}{\mathcal{D}}
\newcommand{\cW}{\mathcal{W}}
\newcommand{\cQ}{\mathcal{Q}}
\newcommand{\dg}{\dagger}
\newcommand{\da}{\dagger}
\newcommand{\eps}{\varepsilon}
\newcommand{\Op}[1]{\hat{#1}}
\newcommand{\ophi}{\Op{\varphi}}
\newcommand{\opi}{\Op{\pi}}
\newcommand{\osigma}{\Op{\sigma}}
\newcommand{\oL}{\Op{L}}
\newcommand{\oH}{\Op{H}}
\newcommand{\oU}{\Op{U}}

\renewcommand{\tr}{\operatorname{tr}}
\newcommand{\dd}{\mathrm{d}}
\newcommand{\curly}[1]{\ensuremath{\left\{#1\right\}}}

\begin{document}

\title{Work production of quantum rotor engines}

\author{Stella Seah$^1$, Stefan Nimmrichter$^2$, Valerio Scarani$^{1,2}$}
\address{$^1$ Department of Physics, National University of Singapore, 2 Science Drive 3, Singapore 117542, Singapore}
\address{$^2$ Centre for Quantum Technologies, National University of Singapore, 3 Science Drive 2, Singapore 117543, Singapore}

\date{\today}

\begin{abstract}
We study the mechanical performance of quantum rotor heat engines in terms of common notions of work using two prototypical models: a mill driven by the heat flow from a hot to a cold mode, and a piston driven by the alternate heating and cooling of a single working mode. We evaluate the extractable work in terms of ergotropy, the kinetic energy associated to net directed rotation, as well as the intrinsic work based on the exerted torque under autonomous operation, and we compare them to the energy output for the case of an external dissipative load and for externally driven engine cycles. Our results connect work definitions from both physical and information-theoretical perspectives. In particular, we find that apart from signatures of angular momentum quantization, the ergotropy is consistent with the intuitive notion of work in the form of net directed motion. It also agrees with the energy output to an external load or agent under optimal conditions. This sets forth a consistent thermodynamical description of rotating quantum motors, flywheels, and clocks.
\end{abstract}


\section{Introduction}\label{sec:intro}

From heat engines to refrigerators, thermodynamics undoubtedly plays an integral role in our everyday life. With modern technologies being able to reduce things down to the nanoscale where quantum effects may not be negligible, there has been an increased interest in the applicability of thermodynamics in the quantum regime. For the study of heat engines, both autonomous \cite{tonner2005,youssef2010,brunner2012virtual,gilz2013,mari2015quantum,kosloff2016flywheel,alex2017rotor,hardal2017} and non-autonomous \cite{kosloff1984,scully2003,rezek2006,alicki2014,zhang2014,uzdin2016} models have been proposed. Non-autonomous heat engines require external control for work extraction, whereas autonomous engines are self-contained. The notion of work may be well-established in the classical regime, but it is not unanimous when we look into quantum systems~\cite{roncaglia2014,hanggi2016,sai2016review,masahito2017}. This is because, unlike observables such as position and momentum associated to a given \emph{state}, work characterizes a given \emph{process}~\cite{hanggi2007}. In fact, many theoretical studies have looked at work extraction for arbitrary quantum systems~\cite{allah2004work,skrzypczyk2014work,binder2015work,acin2015,niedenzu2017work}.

In this paper, we apply and compare common definitions of work to analyze the thermodynamic performance of autonomous rotor heat engines in the quantum regime. The essential component of these self-contained engine models is a planar rotor degree of freedom representing a piston that is coupled to a given quantum working fluid. Its periodic motion defines a continuous engine cycle, and it accelerates by virtue of the force exerted by the thermally driven working fluid, thus integrating an intrinsic clock and a ``flywheel'' work storage. We demonstrate that, even in the quantum regime of low inertia and noisy operation, the amount of extractable work is stored in the form of angular momentum and captured fairly accurately by the kinetic energy associated to the net directed piston motion---a distinct and intuitive perk of rotor engines. Moreover, we show that one can extract the greater part of it by ``putting the wheels on the ground'', i.e.~attaching a dissipative load to the rotor.

We first establish the generic rotor engine setting in Section \ref{sec2}, and review the relevant notions of work. In Section \ref{sec:models}, we then introduce two prototypical engine models and examine their work performance. 
The first model, based on a modified three-mode interaction from previous works \cite{linden2010fridge,levy2012fridge, brunner2012virtual,gilz2013,brask2015,mitchisonFridge2015,kosloff2016flywheel}, realizes a quantum mill, or ``flywheel engine'': rotation is driven by a directed heat flow between two resonantly coupled qubits (or harmonic modes), each thermalizing with its local (hot or cold) environment.
The second model, closer in spirit to a real-life piston engine, is comprised of a single-qubit working fluid that simultaneously interacts with a hot and a cold bath \cite{alex2017rotor}. The working fluid heats up and cools down as a function of the rotor angle, resulting in alternating strokes that accelerate the piston and subject it to backaction noise.
In Section \ref{sec:steady}, we then consider stationary engine operation in the presence of an external dissipative load attached to the rotor. There, the work output will be measured in terms of the dissipated energy, according to a recently developed damping-diffusion model for the orientation state \cite{benjamin2017}. 
Finally, in Section \ref{sec:externaldriving}, we compare the extractable work predicted by these dynamical models with that of their externally controlled, non-autonomous counterparts where the piston coordinate is made to rotate at a controlled angular frequency.

\section{Work on a rotor piston}
\label{sec2}

We consider a generic quantum rotor engine setting comprised of a single planar rotor degree of freedom for the engine piston and one or few bound quantum degrees of freedom for the working fluid. 
The piston is characterized by its moment of inertia $I$, and by the canonical operators $(\ophi, \oL)$ for the angle and angular momentum, with $\oL = \sum_{\ell=-\infty}^\infty \hbar\ell \ket{\ell} \bra{\ell} $ and $\braket{\varphi}{\ell} = e^{i\ell\varphi}/\sqrt{2\pi}$. 

We assume rigid coupling between the piston position and the working fluid, described by an angle-dependent interaction $\oH_{\rm int} (\ophi)$ mediating between the free Hamiltonian $\oH_0$ of the fluid and the kinetic energy $\oL^2/2I$ of the piston. 
The corresponding torque will be denoted by
\begin{equation}\label{eq:torque}
    \Op{F}\left(\ophi\right) = -\frac{\partial \oH_\mathrm{int}}{\partial\ophi}.
\end{equation}
The working fluid, or separate parts thereof, is in contact with a hot and a cold thermal reservoir, as described by two Lindblad superoperators $\cL_{\rm H}^{\ophi} $ and $\cL_{\rm C}^{\ophi}$. In general, we allow them to depend on the piston coordinate $\ophi$, in which case they predict an effective measurement backaction on the piston in the form of angular momentum diffusion. For fixed $\varphi$-values, however, each superoperator should predict thermalization of the free working fluid to a Gibbs state of temperature $T_{\rm H,C}$, i.e.~$\cL_{\rm H,C}^{\varphi} e^{-\oH_0/k_{\rm B} T_{\rm H,C}} = 0$. The overall master equation for the autonomous quantum engine reads as
\begin{equation}\label{eq:genericME}
\dot{\rho} = -\frac{i}{\hbar} \left[ \oH_0 + \frac{\oL^2}{2I} + \oH_{\rm int} (\ophi), \rho \right] + \cL_{\rm H}^{\ophi} \rho + \cL_{\rm C}^{\ophi} \rho.
\end{equation}
Note that, to ensure validity of this equation, we consider typical  transition frequencies of the free Hamiltonian $\oH_0$ much greater than the inverse correlation times of the thermal environments, which in turn exceed the piston-induced modulation of the working fluid associated to $\oH_{\rm int} (\ophi)$ \cite{alicki1979work}. 
In Section \ref{sec:models}, we will study two paradigmatic engine models described by master equations of the above type, see Fig.~\ref{fig:sketches}.

Contrary to textbook examples, an autonomous rotor engine described by \eqref{eq:genericME} will not evolve on predefined engine cycles synchronized to the pointer of an external clock. That role will instead be played by the engine's piston and its dynamical variable $\ophi$.
We now introduce intrinsic figures of merit for the work that accumulates in the isolated piston generated by the isolated engine, which we will compare later with the actual work output to an attached load or external control agent.

\subsection{Intrinsic work definitions}\label{sec:intrinsicwork}

For an intrinsic definition of mechanical output, we must look at the net energy transfer from the working fluid, which continuously receives and dispenses heat, to the rotor piston. The latter acts as an integrated ``flywheel'' battery, and any change of its kinetic energy is entirely caused by the thermally driven dynamics of the working fluid,
\begin{equation}\label{eq:workKE}
    \dot{\cW}_{\rm kin}(t) = \frac{\dd}{\dd t} \frac{\mean{\oL^2}}{2I} = \dot{\cW}_{\rm int} (t) + \frac{1}{2I} \mean{\left( \cL_{\rm H}^{\ophi} + \cL_{\rm C}^{\ophi}\right)^\da \oL^2 }. 
\end{equation}
This power is comprised of two parts. The first one describes how the kinetic energy changes in response to the torque \eqref{eq:torque} exerted by the working fluid, 
\begin{equation}\label{eq:intrinsicwork}
   \dot{\cW}_{\rm int}(t) = \frac{\mean{ \left\{ \oL, \Op{F} (\ophi) \right\} } }{2I} .
\end{equation}
It is the quantum version of the work done by a classical force on the piston over time, $\delta W = F(\varphi) \dd \varphi$ or $\dot{W} = F(\varphi) L/I$. 

The second part in \eqref{eq:workKE} reflects the systematic effect and the backaction noise that comes directly from thermalization. It is only present if the Lindblad dissipators representing the thermal contact to the baths explicitly depend on $\ophi$; the incoherent energy exchange between the working fluid and the environment then constitutes an effective nonselective measurement channel for the piston position that results in momentum diffusion \cite{alex2017rotor}.

The powers \eqref{eq:workKE} or \eqref{eq:intrinsicwork} describe, with or without backaction, the rate at which the working fluid stores kinetic energy in the piston. However, not all of this energy is useful in the sense that it could be extracted and consumed to, say, lift a weight in a deterministic manner. In the worst case, the working fluid would just heat up the piston by exerting a random out-of-sync force, and the resulting kinetic energy would be entirely disordered (or \emph{passive}) and could not fuel any useful deterministic task. 

Classical intuition tells us that net clockwise or anti-clockwise motion reflects the amount of useful energy stored in the rotor piston, $\cW_{\rm net} (t) = \la \oL\ra^2/2I$. This motivates the heuristic definition of useful work in terms of the net kinetic energy that accumulates in the piston degree of freedom as the engine starts moving,
\begin{equation}\label{eq:worknetKE}
    \dot{\cW}_{\rm net} =\frac{\dd}{\dd t} \frac{\la \oL\ra^2}{2I} = \frac{\la\oL\ra}{I} \mean{ \Op{F} (\ophi) + \left( \cL_{\rm H}^{\ophi} + \cL_{\rm C}^{\ophi}\right)^\da \oL }.
\end{equation}
It is upper-bounded by \eqref{eq:workKE} and, in the absence of backaction, by \eqref{eq:intrinsicwork}. 

In Section \ref{sec:externaldriving}, we will replace the dynamical piston degree of freedom by a fixed external time dependence, $(\ophi,\oL) \to (\omega t, I\omega)$. The expressions \eqref{eq:workKE}, \eqref{eq:intrinsicwork}, and \eqref{eq:worknetKE} will then all reduce to the same conventional work definition \eqref{eq:work} for time-dependent systems, suggesting that the subtle differences in the definition of work arise in the autonomous scenario.

\subsection{Maximum Extractable Work: Ergotropy}\label{sec:ergotropy}

In the absence of an external agent that would continuously extract work during operation, the mechanical output of the working fluid accumulates in the accelerating piston. The amount of work the external agent could extract after a given time is upper-bounded by the \emph{ergotropy} \cite{allah2004work,goold2016review}. It is defined for a system Hamiltonian $\oH$ and state $\rho$ as the greatest amount of energy that can be removed by means of a unitary (representing an ideal cyclic process without dissipation),
\begin{equation}\label{eq:ergotropy}
\cW_\mathrm{erg} = \max_{\oU} \curly{ \tr \! \left[\oH (\rho - \oU \rho \oU^\da) \right] } = \tr ( \rho \oH)- \tr ( \opi \oH).
\end{equation}
The resulting transformed state $\opi$ is said to be \emph{passive} \cite{lenard1978work,pusz1978passive,niedenzu2017work}, i.e.~its energy content cannot be reduced further by means of another unitary. This implies that $\opi$ must be diagonal in the energy eigenbasis $\{ \ket{\eps_n} \}$ and its eigenvalues decrease with growing energy $\eps_{n+1} \geq \eps_n$. Note that the ergotropy is always upper-bounded by the free energy difference between $\rho$ and a Gibbs state of the same entropy \cite{allah2004work}.

Given the spectral decomposition $\rho = \sum_n p_n \ket{v_n} \bra{v_n} $ in descending order, $p_{n+1} \leq p_n$, we get explicitly $\opi = \sum_n p_n \ket{\eps_n} \bra{\eps_n}$ for the passive state, and $\oU = \sum_n \ket{\eps_n} \bra{v_n}$ for the corresponding cyclic unitary. Hence, 
\begin{equation}\label{eq:ergotropyfinal}
\cW_\mathrm{erg} = \sum_{n,m} \eps_n p_m \left |  \braket{\eps_n}{v_m}\right |^2 - \sum_{n}\eps_n p_n.
\end{equation}
Here we define ergotropy with respect to the piston degree of freedom, assuming that, naturally, any external load or work extraction process can only access the energy stored in the piston. 
We thus consider the reduced rotor state for $\rho$ and $\oH = \oL^2/2I$.
This way, we do not count in hypothetical work associated to correlations (and free energy differences) between piston and working fluid that could be extracted by means of global unitaries. 

\section{Rotor Heat Engine Models}\label{sec:models}

We now study the work performance of the two most elementary prototypical models for autonomous rotor heat engines. They will realize continuous engine cycles rather than clearly separated heat and work strokes. Both models predict a self-acceleration of the rotor piston driven solely by the temperature bias between two conventional thermal reservoirs without external driving. 

All the following numerical results were obtained from a simulation of the underlying engine master equations on a truncated rotor Hilbert space using the QuTiP package \cite{johansson2013qutip}. We assume that the piston always starts from its ground state $\ket{\ell=0}$ at rest, avoiding external energy input at the initialization stage. When computing the time derivative of the ergotropy \eqref{eq:ergotropyfinal}, we remove numerical fluctuations with help of a variational scheme for numerical differentiation \cite{chartrand2011noisydiff}.

\subsection{Quantum Mill}\label{sec:newmodel}

\begin{figure}
\centerline{\includegraphics{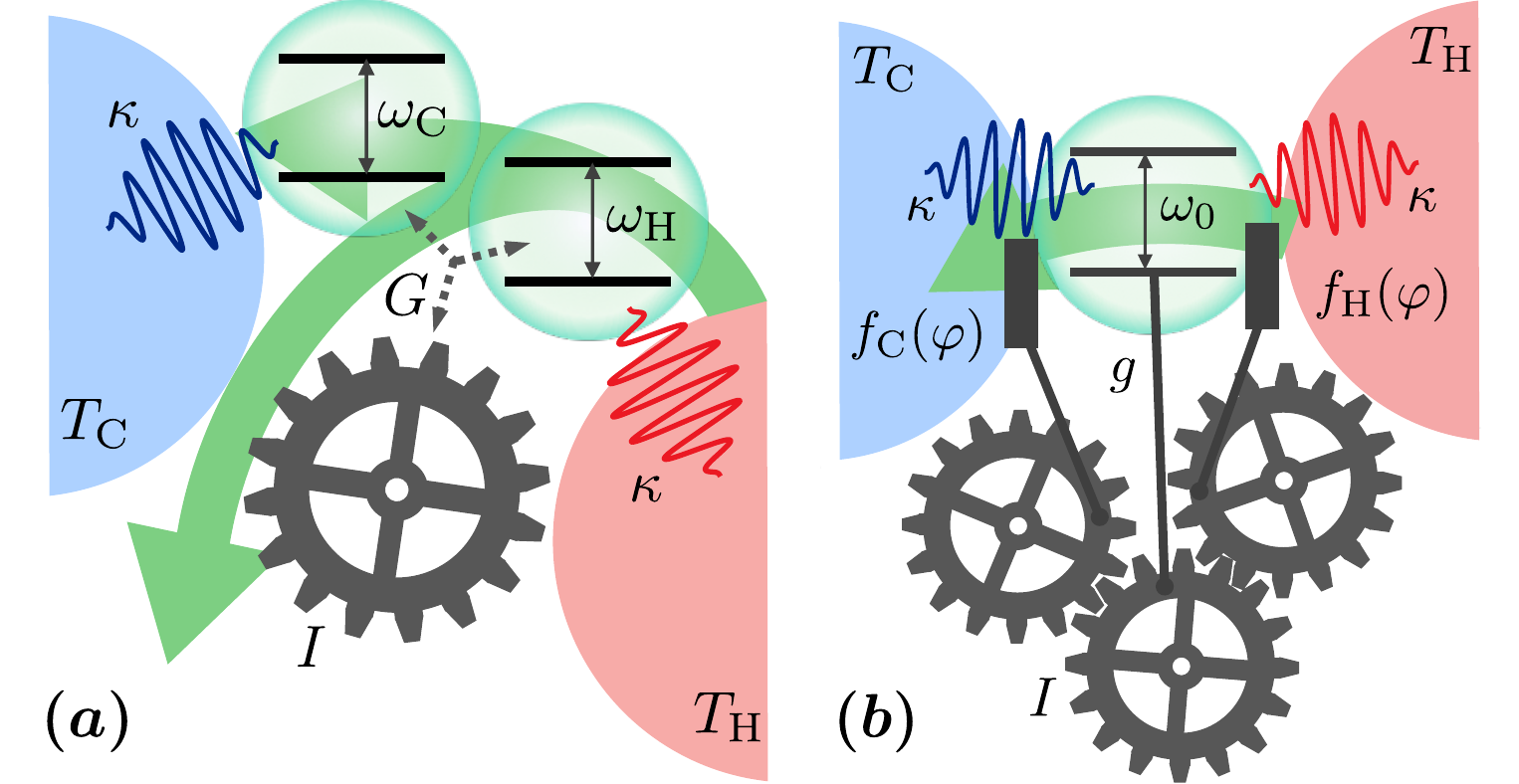}}
\caption{(a) Schematic diagram of the quantum mill. Two qubits with energy gaps $\hbar\omega_{\rm C}$ and $\hbar\omega_{\rm H}$ (here $\omega_{\rm C}=\omega_{\rm H}$) are separately in thermal contact with heat baths of temperatures $T_{\rm H} > T_{\rm C}$ at thermalization rates $\kappa$. The qubits exchange excitations through a rotor with moment of inertia $I$ at the rate $G$, thereby transferring angular momentum quanta that gradually accelerate the rotor in the direction of the heat flow. 
(b) Quantum piston engine. A qubit with energy gap $\hbar\omega_0$ is in thermal contact with two heat baths whose thermalization rates are modulated as functions $f_{\rm H,C}$ of the rotor angle $\varphi \in [0,2\pi)$. The rotor also acts as a piston that experiences a downward push (towards $\varphi = \pi$) by the excited qubit at coupling rate $g$. It will spin up clockwise in a setting where the hot bath coupling dominates for $\varphi < \pi$ and the cold one for $\varphi > \pi$.}\label{fig:sketches}
\end{figure}

In the most basic engine scheme, we employ the rotor as a quantum mill wheel that is accelerated by a steady directed heat flow from a hot to a cold thermal reservoir, see Fig.~\ref{fig:sketches}(a). The working fluid that accommodates the heat flow and exerts the corresponding pressure on the rotor consists of two qubits, $\oH_0 = \hbar\omega_{\rm H} \osigma^{+}_{\rm H} \osigma^{-}_{\rm H} + \hbar\omega_{\rm C} \osigma^{+}_{\rm C} \osigma^{-}_{\rm C}$. Each shall be coupled to its own harmonic oscillator bath, hot (H) or cold (C), assuming standard weak-coupling dissipators \cite{gardiner1985, Breuer2002, carmichael2003, gardiner2004noise} for the thermalization terms in the engine master equation \eqref{eq:genericME},
\begin{equation} \label{eq:Ltherm}
\cL_{j} \, \rho = \kappa \bar{n}_{j} \cD \left[ \osigma^{-}_{j} \right] \rho + \kappa  (\bar{n}_{j} + 1) \cD \left[ \osigma^{+}_{j} \right] \rho.
\end{equation}
Here, $\cD[\Op{O}]\rho = \Op{O}\rho \Op{O}^\da - \{ \Op{O}^\da \Op{O}, \rho \}/2$, $\kappa$ denotes the thermalization rate parameter for both qubits, and $\bar{n}_{j}$ the mean occupation number of oscillators at frequency $\omega_j$ and temperature $T_j$ in bath $j= {\rm C,H}$. There is no backaction-inducing angle dependence in the dissipators.

For the heat exchange between the qubits pushing the rotor, we consider the interaction Hamiltonian \cite{gilz2013}
\begin{equation}\label{eq:Hintflywheel}
\oH_\mathrm{int} (\ophi) =  \hbar G \left( \osigma^-_{\rm H} \osigma^+_{\rm C} e^{i\ophi} + \osigma^+_{\rm H} \osigma^-_{\rm C} e^{-i\ophi}\right),
\end{equation}
which describes a positive or negative angular momentum kick for each excitation transfer from the hot to the cold qubit or vice versa. 
The heat exchange at rate $G$ thus makes the rotor perform an incoherent random walk in angular momentum steps of $\pm \hbar$, and the temperature bias induces a momentum bias that accumulates in the rotor. Once the latter spins at an average frequency $\la \oL \ra /I$, it will act as an effective detuning between the qubits and eventually stifle the heat exchange as soon as the frequency exceeds the qubit linewidth, $\la \oL \ra /I \gtrsim \kappa$. 
For simplicity, we consider the resonant case $\omega_{\rm H} = \omega_{\rm C}$, which allows us to ignore the free Hamiltonian $\oH_0$.

In the regime of fast thermalization, $\kappa \gg G, \la \oL \ra /I$, we can eliminate the qubits by assuming they are always in thermal equilibrium with their respective baths. The approximate master equation for the reduced state $\rho_{\rm R}$ of the rotor (see \ref{sec:derivationME} for a derivation) is
\begin{equation}\label{eq:MErot}
\dot{\rho}_{\rm R} \approx -\frac{i}{\hbar}\left[\frac{\oL^2}{2I},\rho_{\rm R} \right] + \xi \bar{n}_{\rm C}(\bar{n}_{\rm H}+1) \cD\left[e^{-i\ophi}\right]\rho_{\rm R} + \xi \bar{n}_{\rm H} (\bar{n}_{\rm C}+1) \cD\left[e^{i\ophi}\right]\rho_{\rm R}.
\end{equation}
The rotor is effectively pushed by incoherent angular momentum kicks into positive or negative direction, depending on the temperature difference. 
The kick rate decreases the faster the qubits thermalize,
\begin{equation}\label{eq:kickrate}
\xi = \frac{2G^2}{\kappa(\bar{n}_{\rm H}+\bar{n}_{\rm C}+1)(2\bar{n}_{\rm H}+1)(2\bar{n}_{\rm C}+1)} \ll G.
\end{equation} 
The mean and the variance of the angular momentum grow linearly with time, 
\begin{equation}\label{eq:meanLgainA}
\frac{\dd  }{\dd t} \la \oL \ra \approx \hbar \xi (\bar{n}_{\rm H} - \bar{n}_{\rm C} ), \quad
\frac{\dd  }{\dd t} \Delta L^2 \approx \hbar^2 \xi (2 \bar{n}_{\rm H} \bar{n}_{\rm C} + \bar{n}_{\rm H} + \bar{n}_{\rm C}),
\end{equation}
which implies that the signal-to-noise ratio $\la \oL\ra / \Delta L$ will improve like $\sqrt{t}$. 
Starting from the rotor ground state, good signal-to-noise beyond unity is reached once $\xi t > (2 \bar{n}_{\rm H} \bar{n}_{\rm C} + \bar{n}_{\rm H} + \bar{n}_{\rm C})/(\bar{n}_{\rm H} - \bar{n}_{\rm C})^2$, provided that $\la \oL \ra / I \ll \kappa$. 

\begin{figure}
\centerline{\includegraphics{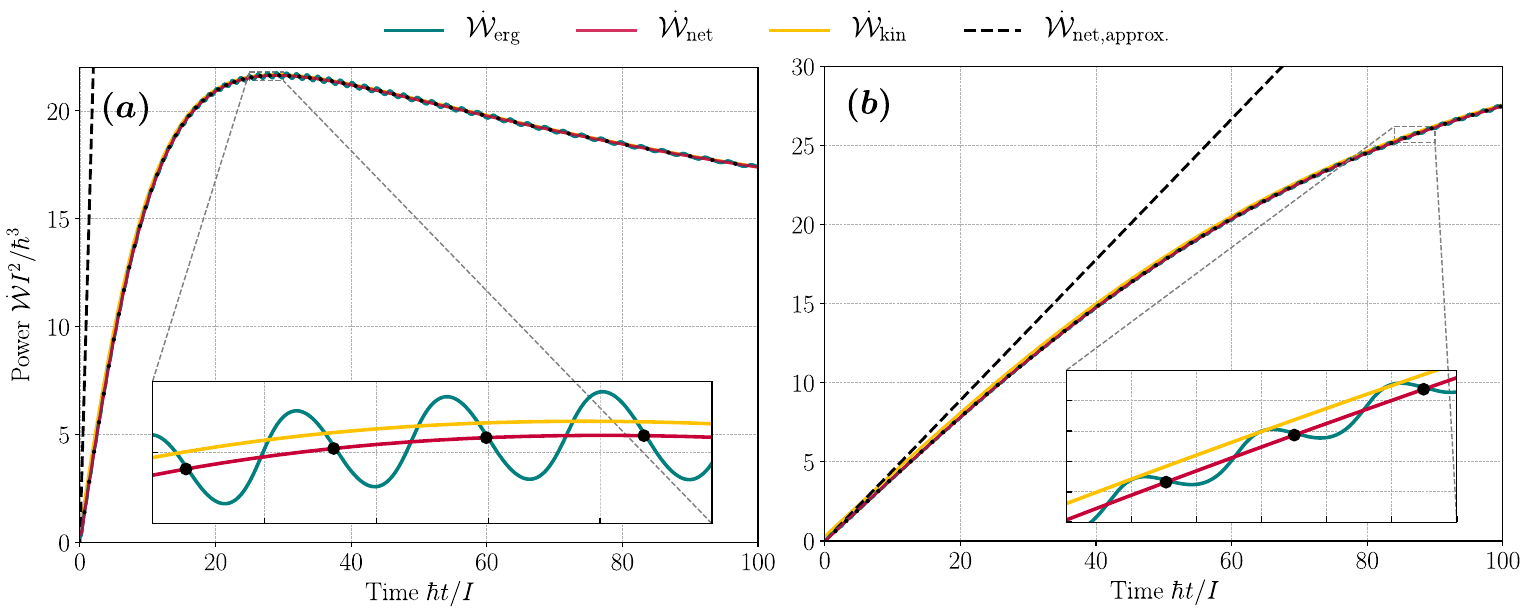}}
\caption{Mechanical power transferred to the rotor under transient, autonomous operation of the quantum mill model. We plot the time derivatives of kinetic energy (yellow), net kinetic energy (red), and ergotropy (green). The dashed line gives the net kinetic energy predicted by the effective model \eqref{eq:MErot}. Markers are placed on the red lines in the magnified details where the momentum $\la \oL \ra$ reaches multiples of $\hbar$. We compare two settings of slow and fast thermalization, (a) $\kappa = 10\hbar/I$ and (b) $50\hbar/I$. The heat exchange rate and the bath temperatures are fixed to $G = 10 \hbar/I$ and $(\bar{n}_\mathrm{C},\bar{n}_\mathrm{H})=(0,1)$. We ran simulations with a rotor Hilbert space truncated to $-50 \leq \ell \leq 250$.}\label{fig:diffworkflywheel}
\end{figure}

The present model is inspired by previous studies based on harmonic or unbounded linear degrees of freedom in place of the rotor \cite{gilz2013,kosloff2016flywheel}. In the working fluid, one may easily substitute harmonic oscillators for qubits by replacing all ladder operators $\osigma_{j}^{\pm}$ with their bosonic equivalents. 
The validity of master equations based on local thermalization channels \eqref{eq:Ltherm} and a simultaneous energy exchange was studied in \cite{rivas2010,levy2014, hofer2017, gonzalez2017}. These models typically fail to give a correct physical description for strong coupling, $G \gg \kappa$. 

An exemplary time evolution of the rates at which work accumulates in the rotor is plotted in Fig.~\ref{fig:diffworkflywheel}(a), comparing the intrinsic mechanical power expressions of Sec.~\ref{sec:intrinsicwork} with the time derivative of the ergotropy \eqref{eq:ergotropyfinal}. With no backaction present, the power \eqref{eq:intrinsicwork} associated to the intrinsic force equals the rate of change \eqref{eq:workKE} in kinetic energy (yellow solid line), 
\begin{equation} \label{eq:intrinsicworkA}
\dot{\cW}_{\rm int}=\dot{\cW}_{\rm kin} = \frac{\hbar G}{I}\mathrm{Im}{\mean{\osigma^-_{\rm H} \osigma^+_{\rm C} \left\{e^{i\ophi},\oL\right\} } }.
\end{equation} 
It overestimates the change \eqref{eq:worknetKE} in net kinetic energy (red) only slightly. Note that the intrinsic work expressions would deviate more in the strong coupling regime, $G \gg \kappa$. Backaction would then enter the game, as the local master equation \eqref{eq:genericME} would no longer be valid and a global master equation should be used instead \cite{levy2014,hofer2017,gonzalez2017}.

The ergotropy change (green) exhibits an oscillatory pattern around its baseline, which coincides with the net kinetic energy (red) curve when the mean angular momentum $\la \oL \ra$ assumes an integer multiple of $\hbar$ (black markers). This is a result of angular momentum discretization and represents the only genuine quantum signature in the otherwise incoherent rotor dynamics.

The approximate master equation \eqref{eq:MErot} predicts a linear increase of the extractable power, $\dot{\cW}_{\rm net} \approx \hbar^2\xi^2(\bar{n}_\mathrm{H}-\bar{n}_\mathrm{C})^2/I $. For small $\kappa$ in Fig.~\ref{fig:diffworkflywheel}(a), the approximate model (dashed line) does not hold right from the start, whereas it agrees with the initial slope of the exact result in the case (b) of larger $\kappa$. There, the model ceases to be valid for greater times when the angular frequency of the rotor is comparable to $\kappa$, viz.~$\la\oL\ra/I > 0.24\kappa $ for $\hbar t/I > 20 $.

\subsection{Quantum Piston Engine} 
\label{sec:oldmodel}


The second rotor engine model we consider is a variation of a previously proposed one \cite{alex2017rotor}, as sketched in Fig.~\ref{fig:sketches}(b). Instead of a harmonic mode, we here employ a single qubit as the working fluid exerting pressure on the rotor piston at the rate $g$ through 
\begin{equation}\label{eq:HS}
	\oH_\mathrm{int} (\ophi)  =\hbar g\,\osigma^+\osigma^- \cos\ophi,
\end{equation}
when excited. The free qubit energy $\oH_0 = \hbar \omega_0 \osigma^+ \osigma^-$ is once again omitted. The rotor angle, on the other hand, controls the thermal coupling of the qubit to a hot and a cold thermal oscillator bath via
\begin{equation} \label{eq:LthermB}
\cL_{j}^{\ophi} \, \rho = \kappa \bar{n}_{j} \cD \left[ \osigma^{-} f_j (\ophi) \right] + \kappa  (\bar{n}_{j} + 1) \cD \left[ \osigma^{+} f_j (\ophi) \right],
\end{equation}
as symbolized by the interlocked gears in Fig.~\ref{fig:sketches}(b). 
This is the quantum version of a continuous Otto motor-like engine, which synchronizes the thermally driven push and the piston position according to (deliberately chosen) coupling functions
\begin{equation} \label{eq:coupling}
f_{\rm H}(\ophi) = \frac{1+\sin\ophi}{2},\quad f_{\rm C}(\ophi) = \frac{1-\sin\ophi}{2}.
\end{equation}
When the piston starts moving down in positive direction from its upper turning point at $\varphi=0$, the working fluid will couple predominantly to the hot bath and thus generate high pressure to accelerate the piston until it passes its lower turning point at $\varphi=\pi$. After that, a predominant coupling to the cold bath removes excitations, and thus the piston moves up faster as a result of reduced pressure. The result is an average net gain of angular momentum at the completion of each cycle. 

Our choice of coupling functions ensures that the engine will eventually start spinning in positive direction autonomously, independent of the initial piston state. In the working regime of fast thermalization, $\kappa \gg \la \oL \ra / I$, we may approximate the excitation probability of the working qubit by means of an angle-dependent average of the hot and cold steady-state values,
\begin{equation}\label{eq:nphi_qubit}
p\left(\varphi\right) = \frac{\bar{n}_{\rm H} f_{\rm H}^2\left(\varphi\right)+\bar{n}_{\rm C} f_{\rm C}^2\left(\varphi\right)}{ \left(2\bar{n}_{\rm H}+1\right)f_{\rm H}^2\left(\varphi\right)+\left(2\bar{n}_{\rm C}+1\right) f_{\rm C}^2\left(\varphi\right)}.
\end{equation}
This probability is bounded by $p(\varphi) < 1/2$, which limits the angular momentum gain, $\dd \la \oL\ra/\dd t \approx \hbar g \la p(\ophi) \sin \ophi \ra < \hbar g/2$. Assuming that the net gain per cycle is small, we may further approximate it by the angle-averaged expression
\begin{equation} \label{eq:meanLgainB}
\frac{\dd }{\dd t} \la \oL \ra \approx \frac{\hbar g}{2\pi} \int_0^{2\pi}  \!\!  \dd \varphi \, p(\varphi) \sin\varphi .
\end{equation}
The qubit-driven piston dynamics is similar to the oscillator-driven one for a suitably rescaled coupling strength $g$. See \ref{sec:qubithocompare} for an explicit comparison at equal maximum gain, which results in a slightly less noisy performance for the qubit case. We will restrict our analysis to this case, which keeps the numerical effort manageable.

\begin{figure}
\centerline{\includegraphics{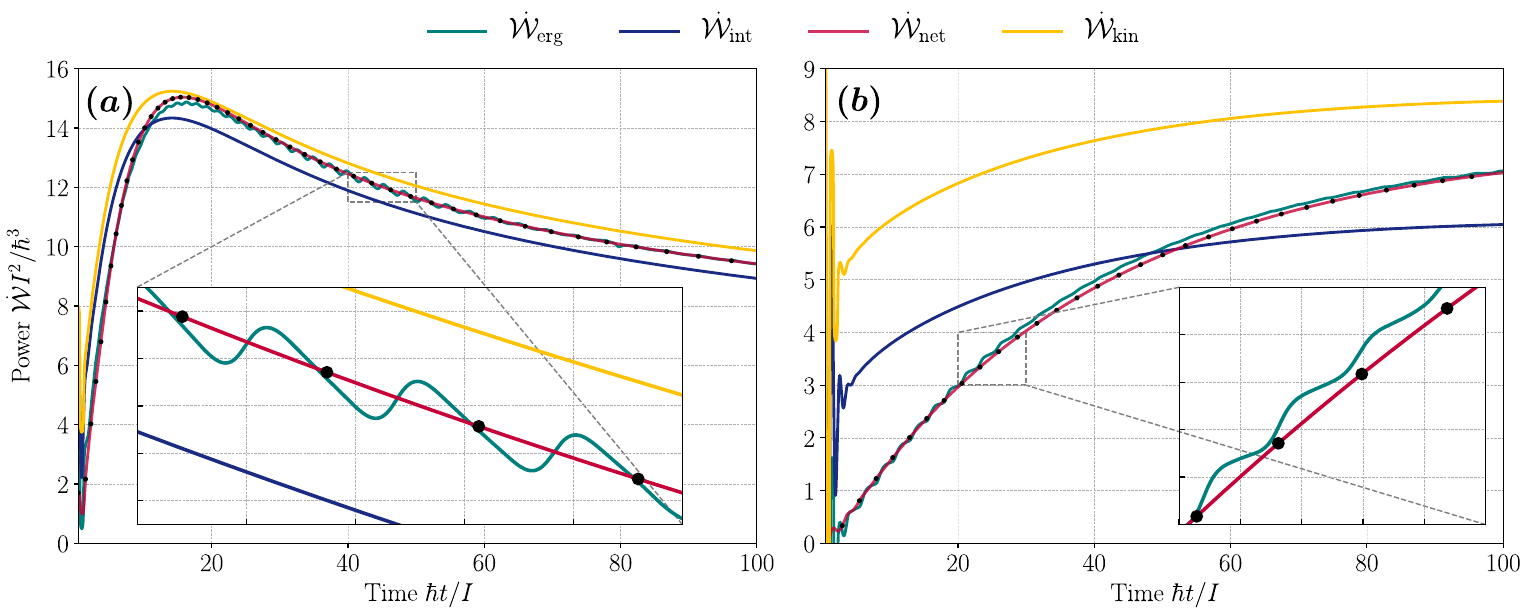}}
\caption{Mechanical power transferred to the rotor under autonomous operation of the quantum piston engine. We compare the power associated to the intrinsic torque (blue) to the growth rates of kinetic energy (yellow), net kinetic energy (red), and ergotropy (green). Markers on the red lines indicate where $\la \oL \ra$ takes multiples of $\hbar$. (a) depicts a setting comparable to the mill model in Fig.~\ref{fig:diffworkflywheel}(a), $\kappa = 10\hbar/I$, $g = 3\kappa$, and $(\bar{n}_\mathrm{C},\bar{n}_\mathrm{H}) = (0,1)$. For (b), we consider higher temperatures at $(\bar{n}_\mathrm{C},\bar{n}_\mathrm{H})=(1,2)$. The results were obtained with a rotor Hilbert space truncated to $-100 \leq \ell \leq 300$.}\label{fig:diffworkold}
\end{figure}

Figure \ref{fig:diffworkold} shows the intrinsic and extractable work rates as a function of time for two different sets of temperatures: (a) the ideal case of a cold bath temperature that approximately amounts to $\bar{n}_{\rm C}=0$ while $\bar{n}_{\rm H}=1$, and (b) the set $(\bar{n}_{\rm C}, \bar{n}_{\rm H}) = (1,2)$. The latter case exhibits a poorer work performance, as the cold bath can cause unwanted excitations that are detrimental to the angular momentum gain per cycle. It also leads to a more pronounced difference between the intrinsic work definitions and the ergotropy (green). In particular, the kinetic energy (yellow) now clearly overestimates the extractable work, indicating that a significant amount of energy accumulating in the rotor is useless, or passive, heat. Part of the reason lies in the backaction noise that contributes to the kinetic energy increase \eqref{eq:workKE} on top of the torque term,
\begin{equation}\label{eq:intrinsicworkB}
    \dot{\cW}_\mathrm{int} = \frac{\hbar g}{2I}\mean{ \osigma^+\osigma^- \curly{\sin(\ophi),\oL}}.
\end{equation}
Quite surprisingly, the mechanical power associated to the torque (blue) \emph{underestimates} the increase in ergotropy at later points in time, when the rates drop as a result of rotations that are as fast as the qubit can react, $\la \oL\ra/I \sim \kappa$. 

Apart from small ergotropy modulations due to angular momentum quantization, the net kinetic energy (red) once again captures the amount of extractable work under all circumstances. This indicates that, in both engine models considered here, useful work appears almost entirely in terms of directed motion, while other contributions to ergotropy, stored in the form of additional quantum coherence, are irrelevant.

\subsection{Performance Comparison} \label{sec:performanceComp}

We now compare the two models in terms of their work performance under equal physical conditions. We consider them at the same qubit resonance frequencies $\omega_{\rm H,C}=\omega_0$, thermalization rates $\kappa$, and bath temperatures $T_{\rm H,C}$ such that $\bar{n}_{\rm C} \approx 0$ and $\bar{n}_{\rm H} = 1$ as in Figs.~\ref{fig:diffworkflywheel} and \ref{fig:diffworkold}(a). 

The key difference between the models is the interaction between the rotor and the working fluid, as characterized by the coupling parameters $G$ and $g$ for the mill and the piston engine model, respectively. For a fair comparison, we choose them to match the angular momentum gain, Eqs.~\eqref{eq:meanLgainA} and \eqref{eq:meanLgainB}, in the working regime of fast thermalization, $\kappa \gg G,g,\la \oL\ra / I$. For the present case, this results in $G \approx 0.534 \sqrt{g\kappa}$.

\begin{figure}
   \centerline{\includegraphics{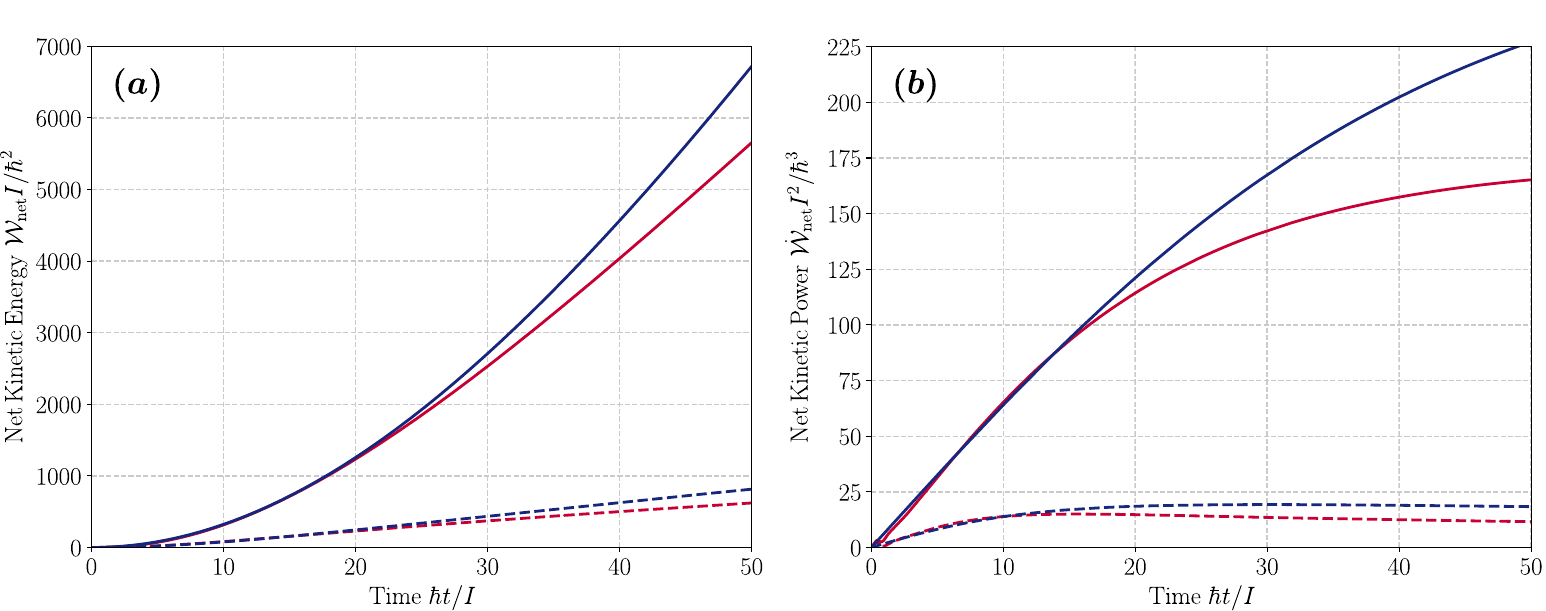}}
\caption{(a) Comparison of the net kinetic energies for the mill (blue) and the piston engine (red) in the cases of fast and slow thermalization, $\kappa = 100\hbar/I$ (solid) and $10\hbar/I$ (dashed). 
(b) Comparison of respective powers, i.e.~time derivatives of the net kinetic energies. The bath occupation numbers are $(\bar{n}_\mathrm{C},\bar{n}_\mathrm{H})=(0,1)$, and the coupling strengths are chosen as $g = 30\hbar/I$ and $G\approx 0.534\sqrt{g\kappa}$ to get the same momentum gain \eqref{eq:meanLgainA} and \eqref{eq:meanLgainB} for both models. For both the solid and dashed lines, we truncated to $ -100\leq \ell \leq 300$ (red) and $-50 \leq \ell \leq 250$ (blue).}\label{fig:netKE2engine}
\end{figure}

Using the net kinetic energy $\cW_\mathrm{net} (t)$ as a figure of merit for the accumulated work, we compare both models in Fig.~\ref{fig:netKE2engine}(a) for the cases of fast (solid lines) and slow (dashed) thermalization. Panel (b) depicts the corresponding power, i.e.~time derivative $\dot{\cW}_\mathrm{net} (t)$. As expected, a higher thermalization rate $\kappa$ increases the attainable work. 
We also note that, while both models initially accumulate the same amount of work, the piston engine (red) is quickly outperformed by the mill model (blue). 
We attribute this to the fact that the piston engine couples to both baths simultaneously, as described by the overlapping coupling functions \eqref{eq:coupling}. This constitutes a direct heat leak from the hot to cold bath lowering the engine performance \cite{chen1997,levy2012quantum,uzdin2015heatleak}. 

In both models, we observe that the work accumulation eventually slows down. Indeed, it is not uncommon that the output power of an engine decreases with fast operating cycles \cite{curzon1975efficiency,geva1994,kosloff2002friction, feldmann2003friction,wang2007friction,cakmak2016}, and we have seen in Figs.~\ref{fig:diffworkflywheel} and \ref{fig:diffworkold} that there exists a point of maximum power beyond which the work rate starts to fall. The critical cycle duration is determined by the finite reaction time of the working fluid, here reached when $\la \oL \ra / I \sim \kappa$. At higher rotation frequencies, the working medium can no longer thermalize effectively.

\section{Steady State Work Extraction}
\label{sec:steady}

So far, we have studied the transient engine behavior under autonomous operation, quantifying the amount of useful work (ergotropy) that accumulates in the rotor degree of freedom and can, \emph{in principle}, be extracted from it at a given time. We have also identified the kinetic energy associated to the net directed rotation as a good figure of merit.

\subsection{Dissipative Load}

In practical scenarios, the rotor will be attached to an external load and output its work under steady-state conditions. The load can be modeled, for example, by means of a classical control field in combination with feedback stabilization \cite{kosloff2016flywheel}, or simply by an additional dissipation channel \cite{mari2015quantum}. We focus on the latter option, which represents the theorist's simplistic version of ``putting the wheels on the ground''. 

At steady state, we will expect a balance of stationary flows: Part of the heat influx from the hot bath dissipates as mechanical output in the load, the remainder is exhausted into the cold bath. The ratio of output to input, the efficiency, will depend on the dissipation rate $\gamma$ characterizing the load's power consumption. From the previous transient analysis, we can already anticipate a sweet spot for maximum efficiency: on the one hand, if $\gamma$ is too low the rotor will spin too fast for the working fluid to thermalize with the baths effectively; on the other hand, at too large $\gamma$ the rotor will not be able to gain enough momentum due to its inertia. Optimal output should be achieved when $1/\gamma$ agrees roughly with the time at which the rotor spins as fast as thermalization occurs, $\la \oL \ra /I \sim \kappa$.

We treat the dissipative load at the rotor by means of a newly developed theory for damping-diffusion of orientation degrees of freedom \cite{benjamin2017}. It contributes an additional Lindblad term $\cL_{\rm R}$ to the master equation \eqref{eq:genericME},
\begin{equation}\label{eq:Lindbladfriction}
\mathcal{L}_{\rm R}\rho = \frac{2k_{\rm B}T_{\rm R} I\gamma}{\hbar^2} \left( \cD\left[\cos\hat{\varphi} - \frac{i\hbar\sin\hat{\varphi}\oL}{4k_{\rm B}T_{\rm R} I}\right] \rho + \cD \left[\sin\hat{\varphi}+\frac{i\hbar\cos\hat{\varphi}\oL}{4k_{\rm B}T_{\rm R} I}\right] \rho \right) ,
\end{equation}
which can be viewed as the analog of the Caldeira-Leggett model of linear Brownian motion \cite{caldeira1983path,Breuer2002} for rotors. Accordingly, it describes angular momentum decay at the rate $\gamma$ and thermalization to a Gibbs-like state of temperature $T_{\rm R}$, including a low-temperature correction to ensure complete positivity as in \cite{diosi1993calderia,vacchini2009}. Contrary to linear motion, $\cL_{\rm R}$ consists of \emph{two} dissipators and is expressed in terms of trigonometric functions to account for the $\ophi$-periodicity and the rotor symmetry.

\subsection{Steady State Output and Efficiency}

For the mechanical output power, we consider the rate at which the system dissipates energy into the load,
\begin{equation}\label{eq:workdissipative}
\dot{\cW}_\mathrm{out} = - \tr \curly{ \left[ \frac{\oL^2}{2I} + \oH_{\rm int} (\ophi) \right] \cL_{\rm R} \rho }.
\end{equation}
At steady state and for appropriate values of the load parameters $\gamma, T_{\rm R}$, this will be a positive number balancing the net energy flow from the hot and cold baths. For the two engine models studied in Sec.~\ref{sec:newmodel} and \ref{sec:oldmodel}, we get 
\begin{equation}\label{eq:dissipativeterms}
\dot{\cW}_{\rm out} = \gamma\mean{\frac{\oL^2}{I} - k_{\rm B} T_{\rm R} - \frac{\hbar^2}{ 16 I k_{\rm B}T_{\rm R}} \left[ \frac{\oL^2}{I} - \oH_{\rm int} (\ophi) \right] } .
\end{equation}
The first two terms reflect the classical equipartition theorem for a free rotor at steady state. Indeed, one recovers $\la \oL^2/2I\ra \approx k_{\rm B} T_{\rm R}/2$ from $\dot{\cW}_{\rm out}=0$ at temperatures $k_{\rm B} T_{\rm R} \gg \hbar^2 / I$ where the thermal distribution extends over many quanta of angular momentum. The low-temperature corrections represented by the remaining terms (and diverging like $1/T_{\rm R}$) can then be neglected. 

\begin{figure}
\centerline{\includegraphics{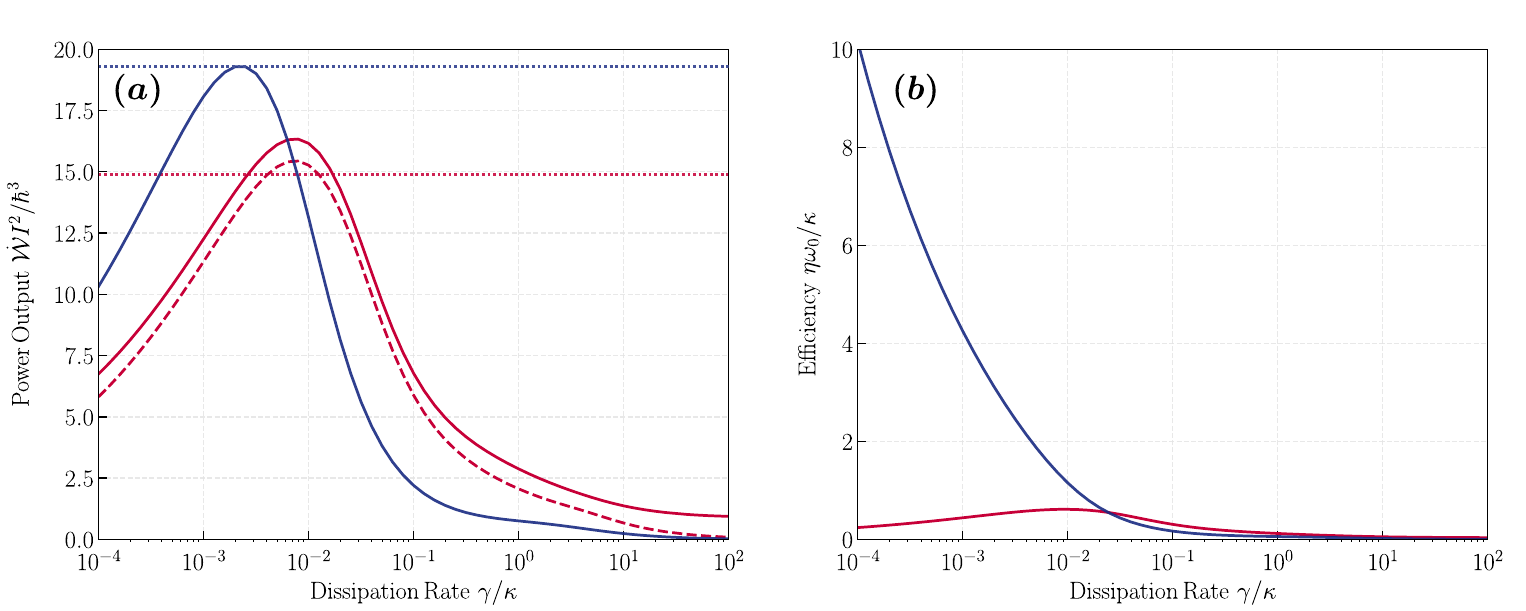}}
\caption{(a) Steady-state output power to load as a function of the dissipation rate $\gamma$ for the mill (solid blue) and the piston engine (solid red). We also plot the power $\dot{\mathcal{W}}_\mathrm{int}$ of the intrinsic torque on the piston (dashed). The dotted lines mark the greatest time derivatives of ergotropy obtained without load for the settings of low $\kappa$ in Fig.~\ref{fig:netKE2engine}.
(b) Efficiency in units of thermalization rate $\kappa$ over qubit frequency, $\kappa/\omega_0 \ll 1$.}\label{fig:friction}
\end{figure}

For the driven engine rotor at steady state, $\dot{\cW}_{\rm out} > 0$ reflects the mechanical output power, provided that the finite-temperature load does not inject appreciable power on its own, i.e.~$\gamma k_{\rm B} T_{\rm R}$ is small compared to the input from the working fluid. 
Figure \ref{fig:friction}(a) shows the steady-state output power as a function of $\gamma$ for the mill (blue) and the piston engine model (red), using the same set of comparable parameters at $\kappa = 10\hbar/I$ as in Fig.~\ref{fig:netKE2engine}. We fix $k_{\rm B} T_{\rm R} = 10\hbar^2/I$ to avoid low-temperature corrections. 
Both models predict a maximum output power at small dissipation rates compared to $\kappa$ and a vanishing output for ``heavy'' loads of large $\gamma$. 

The steady-state efficiencies $\eta = \dot{\cW}_{\rm out}/\dot{\cQ}_{\rm in}$ for the mill and the piston are shown in Fig.~\ref{fig:friction}(b). For this, we approximate the heat input power as $\dot{\cQ}_{\rm in} \approx \tr \{ \oH_0 \cL_{\rm H} \rho \}$ in the weak-coupling regime considered here, $g,G \ll \omega_0$. We use the ratio of the thermal linewidth over the qubit frequency, $\kappa/\omega_0 \ll 1$, as a common reference unit for the efficiency.

In the piston model, maximum output power and efficiency is achieved around $\gamma \sim 10^{-2}\kappa$, and the corresponding time scale $1/\gamma$ matches roughly the time at which the piston rotates as fast as the qubit thermalizes and at which the extractable power in Fig.~\ref{fig:diffworkold}(a) is greatest. The latter is also represented by the red dotted line in Fig.~\ref{fig:friction}(a), which slightly underestimates the maximum output to the load. 

The mill model, on the other hand, achieves its highest output power to the load at smaller dissipation rates and faster rotations, while the efficiency keeps growing as $\gamma$ decreases. Nevertheless, the maximum power matches the greatest value for the time derivative of ergotropy (blue dotted line) that would be obtained without load. This demonstrates that the ergotropy evaluated under autonomous engine operation captures well the extractable work for a dissipative load, but an efficient extraction crucially depends on the load's pull.

In terms of efficiency, the mill clearly outperforms the piston, which we attribute again to the presence of a direct heat leak in the piston model. The more effective use of heat input in the mill model allows the efficiency to grow with increasing steady-state rotation speed (or decreasing $\gamma$) beyond the point of maximum output power. Indeed, and contrary to the piston model, as the mill accelerates, its angular frequency $\Omega = \la \oL \ra/I$ acts as an effective detuning between the two qubits, which suppresses the heat flow. Once they get driven out of resonance to the point where $\Omega > \kappa$, both the net heat exchange and steady-state angular momentum gain are then expected to reduce like $\Omega^{-2}$, as determined from $\oH_{\rm int} (\Omega t)$. This implies that the mechanical output power, i.e.~the rate of change in rotational energy, reduces in proportion to $\Omega^{-1}$, which causes the efficiency to grow. We note however that our master equation model assumes weak thermal bath coupling at the qubit resonance frequency $\omega_0$ and is valid only when $\Omega \ll \omega_0$.

Finally, notice the small residual output power for the piston engine model at very high dissipation rates in Fig.~\ref{fig:friction}(a). In this limit, the load is effectively jamming the piston, and no useful motion is generated. However, the working qubit keeps exchanging excitations with the two thermal baths via the angle-dependent dissipators \eqref{eq:LthermB}, which leads to backaction heating of the rotor \cite{alex2017rotor}. 
In general, its kinetic energy balance has three contributions: 
the intrinsic work power \eqref{eq:intrinsicworkB}, backaction-induced diffusion, and the output \eqref{eq:dissipativeterms} to the load. Omitting the low-temperature corrections, 
\begin{equation}
\frac{\dd}{\dd t } \mean{\frac{\oL^2}{2I}} \approx \dot{\cW}_{\rm int} - \dot{\cW}_{\rm out} + \frac{\hbar^2 \kappa}{4I}\sum_j \mean{(\bar{n}_{j}+\osigma^+ \osigma^-)\left[ f_j'(\ophi) \right]^2},
\end{equation}
which vanishes at steady state. The backaction-induced heating term 
on the right
contributes to the output power and thus constitutes a direct heat leak from the thermal baths to the load. 
It is more significant the higher the bath temperatures.
For comparison, we plot the intrinsic work power $\dot{\cW}_{\rm int}$ under load (red dashed) in Fig.~\ref{fig:friction}(a). It does not give the residual offset and its maximum is closer to the greatest extractable power without load. The discrepancy is minor for the temperatures considered here, and only relevant for the piston model, but it calls attention to the fact that a dissipative load in general does not distinguish between useful work and passive, entropic energy. 

\subsection{Off-Resonant Performance Gain for the Quantum Mill}

So far, we have conveniently assumed that the working medium in the mill model comprises two resonant qubits, $\omega_{\rm H}=\omega_{\rm C} = \omega_0$.  
However, considering the observed interplay between rotation-induced detuning, heat exchange, and momentum gain, one may increase the mill's work performance by detuning the two qubits. Indeed, the acceleration at a given point in time is best when the qubit frequency difference $\Delta = \omega_{\rm H} - \omega_{\rm C}$ is positive and matches the effective detuning $\Omega = \la \oL \ra /I$, and optimal gain would be achieved at all times by adjusting $\Delta$ as the mill wheel spins up. At a fixed $\Delta>0$, on the other hand, the initial acceleration from rest would be suppressed and the mill would take longer to reach the rotation speed of maximum gain.

\begin{figure}
\centerline{\includegraphics{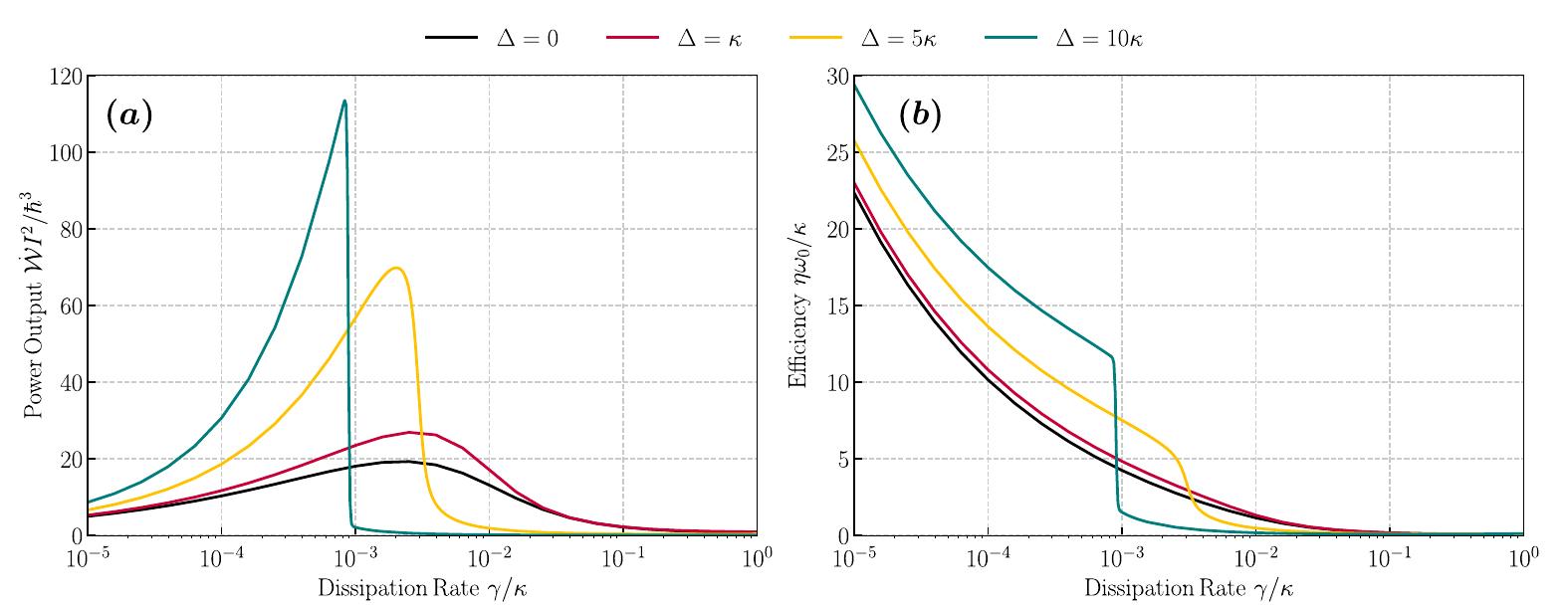}}
\caption{(a) Steady-state output power to load and (b) efficiency as a function of the dissipation rate $\gamma$ for the mill at various detunings $\Delta = \omega_{\rm H} - \omega_{\rm C}$ between the qubits. We assume $\kappa=10\hbar / I$, $G\approx 0.925\kappa$, and $(\bar{n}_{\rm C},\bar{n}_{\rm H}) = (0,1)$ as in Fig.~\ref{fig:friction}, matching the resonant case ($\Delta=0$). For the efficiency, we neglect the dependence of the heat flows on the detuning as we consider the limit $\omega_0 = (\omega_{\rm H} + \omega_{\rm C})/2 \gg \Delta$.}\label{fig:detuning}
\end{figure}

In Figure~\ref{fig:detuning}, we illustrate the effect of a fixed detuning $\Delta$ for an autonomous mill under load. We plot (a) the steady-state output power and (b) the efficiency for various detunings. They range from zero (black, matching Fig.~\ref{fig:friction}) to ten thermal linewidths $\kappa$ (green). We observe that the maximum possible output power to the dissipative load improves drastically with growing detuning, but this maximum is also shifted towards smaller dissipation rates $\gamma$, i.e.~requires a weaker pull of the load.

The corresponding efficiency in (b) exhibits a step-like critical behavior in the off-resonant limit when the detuning exceeds the natural linewidth of the qubits, $\Delta > \kappa$. The step occurs where the stationary rotation frequency of the mill under load matches the detuning, which facilitates a resonant heat flow between the qubits and thereby enhances the output power. For heavier loads on the right of the step, the mill reaches stationary rotation frequencies $\Omega < \Delta$ that are still too low to enhance the heat exchange. Left of the step, the fast rotation begins to suppress the heat exchange again.

\section{External Driving}\label{sec:externaldriving}

In view of the transient behavior of the autonomous rotor engines studied in Sec.~\ref{sec:models}, and in particular the deteriorating work performance after long times, we might ask to what extent this is related to the finite inertia and the dynamical nature of the rotor. It was already shown that autonomous engines with an idealized (dispersion-free, or infinite-mass) internal quantum clock can in principle perform as well as externally driven ones \cite{malabarba2015}. For clocks realized by finite-dimensional quantum systems, the energetic cost or disturbance could be made exponentially small in the clock dimension \cite{Woods2016}. 
Here we ask how well the rotor degree of freedom can fulfill its role as the engine clock.

To this end, we make a comparison to non-autonomous versions of the models where the rotor is replaced by an ideal clock, $\ophi \to \omega t$ with a fixed external driving frequency $\omega$. 
The underlying master equation \eqref{eq:genericME} then becomes of the type introduced in \cite{alicki1979work}, and the conventional work and heat definition for time dependent systems applies. 
Given the slowly varying system Hamiltonian $\oH (t) = \oH_0 + \oH_{\rm int} (\omega t)$, the cumulative net heat \emph{input} and work \emph{output} in a time interval $t$ read as \cite{alicki1979work,binder2015work,niedenzu2017work}
\begin{eqnarray} \label{eq:work} 
\cQ(t) &=& \int_0^t \!\! \dd t' \, \tr[\dot{\rho}(t') \oH(t')] ,  \\ 
\cW(t) &=& -\int_0^t \!\! \dd t' \, \tr[\rho(t')\dot{\oH}(t')] = \int_0^{t} \!\! \omega \dd t' \mean{ \Op{F} (\omega t') }_{t'}, \nonumber 
\end{eqnarray}
The work output rates for the mill and the piston engine model are given by
\begin{equation}\label{eq:workdependent}
 \dot{\cW}^{\rm M} (t) = 2\hbar\omega G \, \mathrm{Im} \left[ e^{i\omega t} \mean{ \osigma_{\rm H}^- \osigma_{\rm C}^+}_t \right] , \quad \dot{\cW}^{\rm P} (t) = \hbar \omega g \sin\omega t\mean{ \hat{\sigma}^+ \hat{\sigma}^-}_t . 
\end{equation}
The heat term is associated to dissipative interactions with the thermal baths, whereas work is associated to the periodic changes in the control parameter $\varphi = \omega t$. After each engine cycle of period $\tau = 2\pi / \omega$, we find $\cQ(\tau) - \cW(\tau) = \la \oH \ra_{\tau} - \la \oH \ra_0$. 
In the ideal limit $\tau \to \infty$ of quasistatic operation, we expect that the right hand side vanishes as the working fluid returns to its initial state at the end. This is no longer the case in finite-time cycles where the work output gradually deteriorates with decreasing $\tau$. 

\begin{figure}
\centerline{\includegraphics{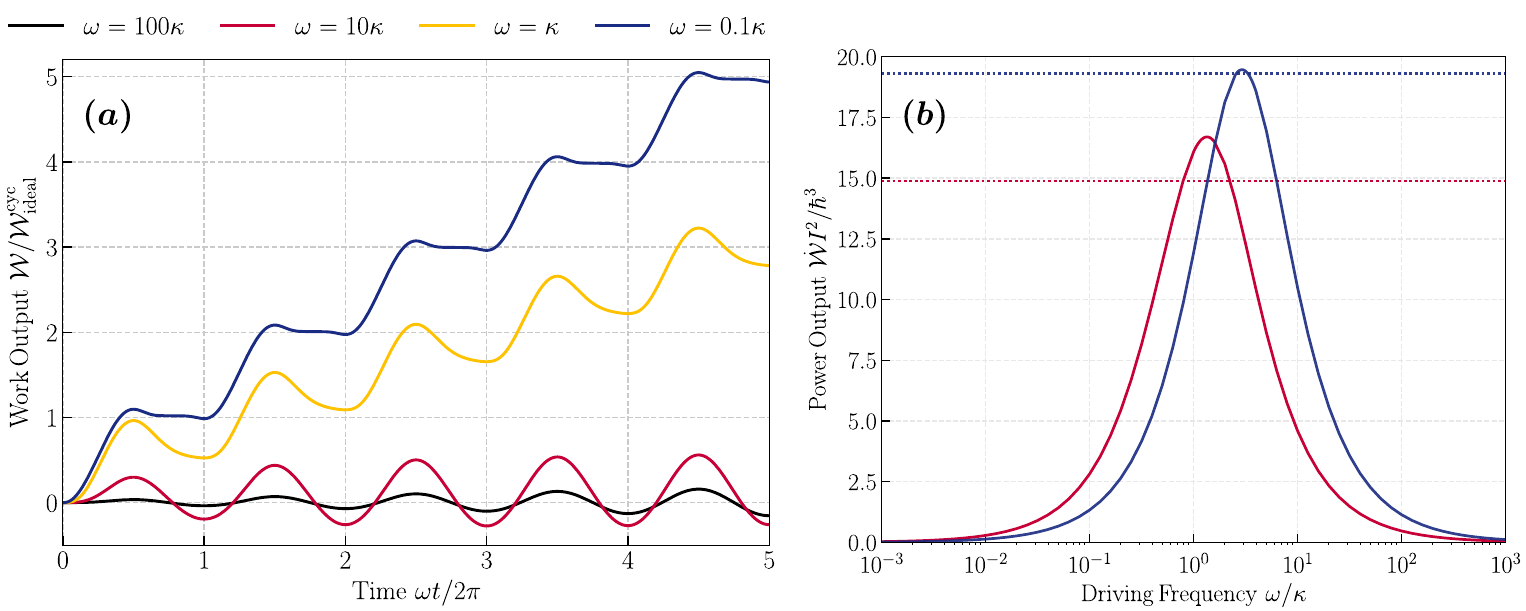}}
\caption{(a) Cumulative work output over time for the non-autonomous piston engine at various driving frequencies $\omega$. Time is given in cycle periods and work in units of the ideal output \eqref{eq:idealwork} per cycle.
(b) Average stationary output power as a function of the driving frequency for the mill (blue) and the piston engine (red). Compare this to the output power under load in Fig.~\ref{fig:friction} using the same settings; also here the dotted lines mark the greatest time derivative of ergotropy without load.}\label{fig:timedep}
\end{figure}

As an instructive example, we consider a simulation of the driven piston engine and plot the accumulated work $\dot{\cW}^{\rm P} (t)$ for various driving frequencies $\omega$ in Fig.~\ref{fig:timedep}(a). The work is given in units of the ideal adiabatic work output per engine cycle $\tau=2\pi/\omega$,
\begin{equation}\label{eq:idealwork}
\mathcal{W}_{\mathrm{ideal}}^\mathrm{cyc} = \hbar g\omega\int_0^{\tau} \!\! \dd t\, p(\omega t) \sin \omega t.
\end{equation}
It assumes that the working qubit thermalizes instantaneously to the angle-dependent excitation probability \eqref{eq:nphi_qubit} yielding an optimal gain \cite{alex2017rotor}. In the case of slow rotation (blue curve), we find that the work output increases in steps of the ideal value. The first half of each step corresponds to the strong push when the qubit couples mostly to the hot bath; the second half exhibits no push as all excitation is dumped in the cold bath. For faster driving exceeding the thermalization rate $\kappa$ (yellow, red, black), we find that the net work output per cycle decreases since the qubit is no more able to follow. 


The cycle-averaged output power $[\cW(t+\tau)-\cW(t)]/\tau$ will eventually reach a stationary value for $t \gg 1/\kappa$. In Figure \ref{fig:timedep}(b), we plot this asymptotic value for different driving frequencies $\omega$, comparing the piston model (red) to the resonant mill model (blue) at comparable parameters. In both cases, maximum output \emph{power} is achieved at $\omega \sim \kappa$, and this maximum also agrees with the greatest time derivative of net kinetic energy (or ergotropy) at autonomous operation without load (dotted lines). In this optimal working regime, autonomous rotor engines are on par with their external clock-driven counterparts.

\section{Conclusion}

We have analyzed and compared the performance of self-contained quantum rotor engine models in terms of intrinsic work storage, ergotropy, and work output to a dissipative load. The models are inspired by the role of rotational degrees of freedom in classical thermal machines: In the first one, the quantum rotor is driven like a mill into directed motion by the stationary heat flow from the hot to the cold end of a working fluid; in the second model, the rotor synchronizes the heating and cooling of a working medium with its pressure on a piston like in an Otto motor. 

In the transient scenario of autonomous engine operation without load, we have shown that the potentially extractable work generated by the engine has a characteristic time dependence achieving optimal power when the rotor spins about as fast as the working medium can react, i.e.~thermalize with the baths. This highlights the rotor's function as a flywheel work storage, even in the noisy quantum regime of operation. Apart from small modulations due to angular momentum quantization, the extractable work (ergotropy) matches closely with the net rotor spin, which consolidates the formal work definition derived from quantum information with physical interpretations.

In the scenario of steady-state operation, we could observe a similar and related behavior with a sweet spot of maximum output power as a function of the coupling rate to an external load attached to the rotor. Here, the load is described in terms of a dissipation model for the orientation state \cite{benjamin2017}. At a properly chosen dissipation rate, the load can extract as much work as predicted by ergotropy under autonomous operation. 

Moreover, by comparing these results to the work output obtained from time-dependent models, where the rotor is replaced by a fixed external driving frequency, we elucidated the rotor's role as an internal engine clock. Once again, the time derivatives of ergotropy and of the energy contained in the net ``ticking'' motion of the autonomous rotor agree well with the highest output powers achieved for external driving frequencies of the order of the engine's thermalization rate. 
Overall, we find a better maximum performance and agreement for the mill model.

Our work demonstrates the benefits of rotor degrees of freedom in the study of quantum thermal machines. Serving both as a flywheel for intermediate work storage and as an internal clock that sets the engine cycle, they facilitate intuitive designs of autonomous quantum engines. The manifestation of work as directed rotation admits an unambiguous assessment of the engine performance in both the quantum and the classical regime.

\emph{Note added in proof:} After submission, we became aware that a similar comparison between ergotropy and realistic work extraction in a different system based on superconducting circuits appeared on the arXiv \cite{Loerch2018}.

\section*{Acknowledgments}

We thank Alexandre Roulet, Benjamin Stickler, and Wolfgang Niedenzu for helpful discussions. This research is supported by the Singapore Ministry of Education through the Academic Research Fund Tier 3 (Grant No. MOE2012-T3-1-009); by the National Research Foundation, Prime Ministers Office, Singapore, through the Competitive Research Programme (Award No. NRF-CRP12-2013-03); and by both above-mentioned sources, under the Research Centres of Excellence programme.

\appendix

\section{Effective master equation for the Quantum Mill model} \label{sec:derivationME}

Here we derive the effective master equation for the reduced rotor state of the quantum mill of Sec.~\ref{sec:newmodel} in the limit of fast thermalization, in analogy to \cite{kosloff2016flywheel}. 
For this, we switch to the interaction picture with respect to the free Hamiltonian $\oH = \oH_0 + \oL^2/2I$ of the two qubits and the rotor,
\begin{equation}\label{eq:MEintflywheel}
\dot{\rho}^{\rm I}=\left(\mathcal{K}_t+\cL_{\rm H} + \cL_{\rm C}\right) \rho^{\rm I}, \quad \mathcal{K}_t\rho^{\rm I} = -\frac{i}{\hbar}\left[\oH^{\rm I}(t),\rho^{\rm I} \right].
\end{equation}
The interaction Hamiltonian is given by
\begin{equation}\label{eq:HintI}
\oH^{\rm I}_t = e^{i\oH t}\oH_\mathrm{int} (\ophi) e^{-i\oH t} = \hbar G \hat{\sigma}^+_{\rm H}\hat{\sigma}^-_{\rm C}e^{-i\hat{\varphi}-i\oL t/I+i\Delta t}  + h.c.,
\end{equation}
with $\Delta= \omega_{\rm H}-\omega_{\rm C} \to 0$ as in the main text. 
We assume that the qubits are initially prepared in Gibbs states, i.e.~in thermal equilibrium with their individual baths, $\rho^{\rm I}(0)=\rho_{\rm H}^{\rm th}{\otimes}\rho_{\rm C}^{\rm th}{\otimes}\rho_{\rm R}^{\rm I}(0)$. Introducing $\cL = \cL_{\rm H} + \cL_{\rm C}$ and using $\cL\rho^{\rm I}(0) = 0$, we can formally integrate \eqref{eq:MEintflywheel} as
\begin{equation}\label{eq:MEimplicitsol}
\rho^{\rm I}(t) = \rho^{\rm I}(0) + \int_0^t \dd s\, e^{\mathcal{L}(t-s)}\mathcal{K}_s\rho^{\rm I}(s).
\end{equation}
Substituting this back into \eqref{eq:MEintflywheel} yields
\begin{equation}
\dot{\rho}^{\rm I} = \mathcal{K}_t\rho^{\rm I}(0) + \left(\mathcal{L} + \mathcal{K}_t \right) \int_0^t \dd s\, e^{\cL s}\mathcal{K}_{t-s} \rho^{\rm I}(t-s).
\end{equation}

\begin{figure}
   \centerline{\includegraphics[width=8cm]{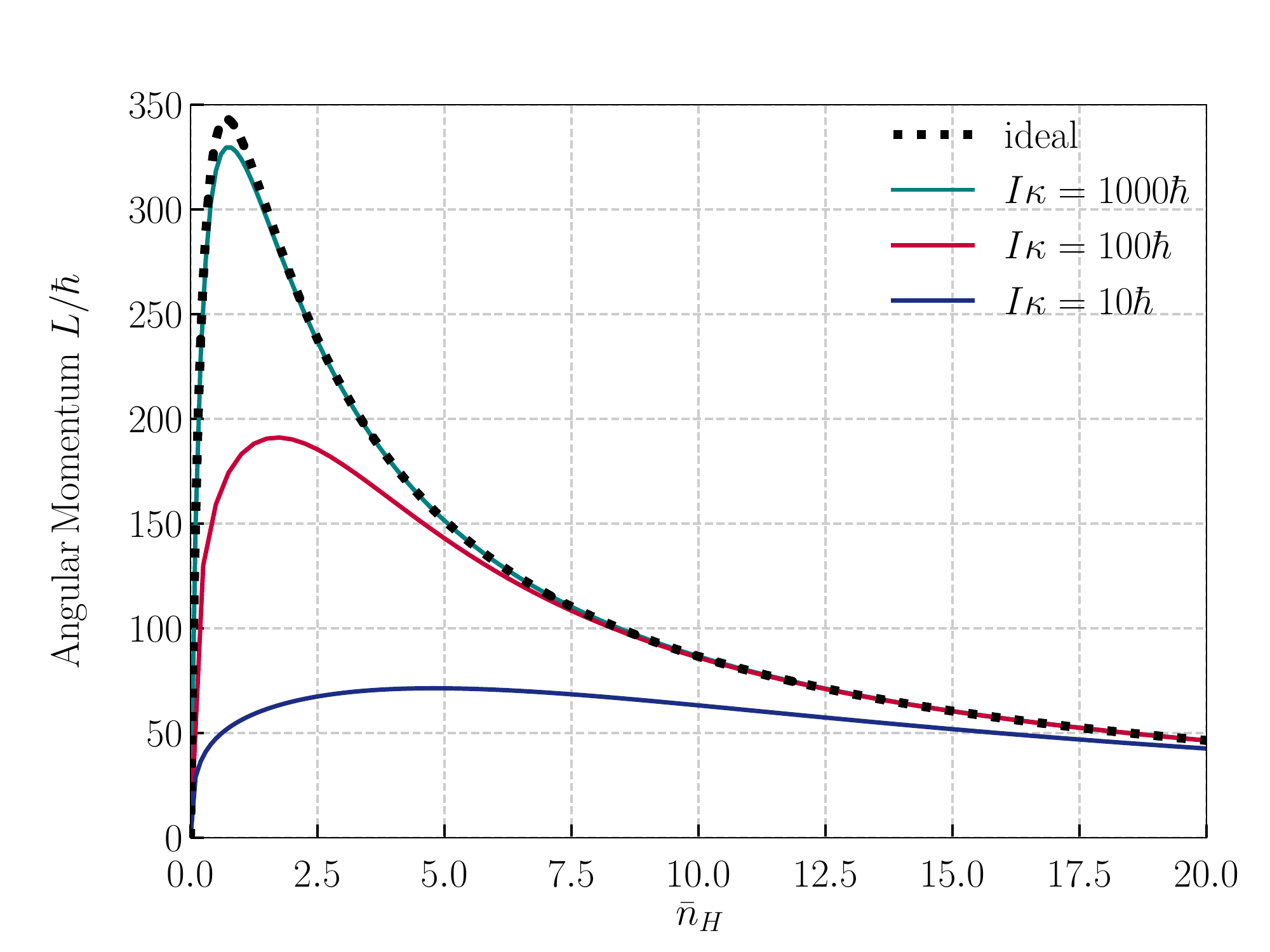}}
\caption{Steady-state value of angular momentum as a function of the hot bath occupation for the mill under dissipative load for different moments of inertia. We assume $\bar{n}_{\rm C}=0$, $G = 10^{-2} \kappa$, and $\gamma=10^{-5}G$. The dashed curve depicts the ideal value $\la \oL \ra_{\rm ss}$ obtained from \eqref{eq:MEfinal} and the dissipative load  term \eqref{eq:Lindbladfriction}. The other curves are obtained by adding the load to the full master equation \eqref{eq:MEintflywheel} of the tripartite state and solving for the steady state numerically.} \label{fig:diffIflywheel}
\end{figure}

In the limit of fast thermalization, we can apply the usual Born-Markov approximations \cite{gardiner1985,Breuer2002, carmichael2003,gardiner2004noise}, 
here $\rho^{\rm I}(t-s) \approx \rho_{H}^{\rm th} {\otimes} \rho_{C}^{\rm th}  {\otimes} \rho_{\rm R}^{\rm I} (t)$ and $\mathcal{K}_{t-s} \approx \mathcal{K}_t$ in the integral and $t\to\infty$ for the upper bound. After taking the partial trace over the qubits with help of the relation $\tr_{\rm HC}\left( e^{\cL s} \mathcal{K}_t \rho_{\rm H}^{\rm th}{\otimes}\rho_{\rm C}^{\rm th}\right)=0$, 
\begin{eqnarray}\label{eq:MEmarkov}
\dot{\rho}_{\rm R}^{\rm I} &=& \tr_{\rm HC}\left(\int_0^\infty \!\! \dd s\, \mathcal{K}_t  e^{\cL s}\mathcal{K}_{t} \rho_{\rm H}^{\rm th} \otimes \rho_{\rm C}^{\rm th} \otimes\rho_{\rm R}^{\rm I} \right) \\
&=& \int_0^\infty \!\! \frac{\dd s}{\hbar^2} \tr_{\rm HC} \curly{ \left[ e^{\cL^\dg s}\oH^{\rm I}_t,\left[\rho_{\rm H}^{\rm th} \otimes \rho_{\rm C}^{\rm th} \otimes \rho_{\rm R}^{\rm I}, \oH^{\rm I}_t \right]\right] }. \nonumber
\end{eqnarray}
The second line is expressed in terms of the conjugate superoperator $\cL^\da$.
One finds $e^{\mathcal{L}^\dg s}\oH^{\rm I}_t = \oH^{\rm I}_t e^{-\kappa s (\bar{n}_{\rm H}+\bar{n}_{\rm C}+1)}$, which leads to
\begin{eqnarray}\label{eq:MEfinal}
\dot{\rho}_{\rm R}^{\rm I} &=& \frac{\tr_{\rm HC} \curly{ \left[\oH^{\rm I}_t , \left[\rho_{\rm H}^{\rm th}\otimes \rho_{\rm C}^{\rm th} \otimes\rho_{\rm R}^{\rm I}, \oH^{\rm I}_t \right]\right]}}{\hbar^2\kappa\left(\bar{n}_{\rm H}+\bar{n}_{\rm C}+1\right)}\\
&=& \xi \bar{n}_{\rm C}(\bar{n}_{\rm H}+1) \cD\left[e^{-i\hat\varphi -i\oL t/I}\right]\rho_{\rm R}^{\rm I} + \xi \bar{n}_{\rm H}(\bar{n}_{\rm C}+1) \cD\left[e^{i\hat\varphi +i\oL t/I}\right]\rho_{\rm R}^{\rm I}, \nonumber
\end{eqnarray}
with $\xi$ defined in \eqref{eq:kickrate}. In the Schr\"odinger picture, the master equation reduces to \eqref{eq:MErot}.
The coefficients in front of the Lindblad dissipators can be interpreted as the rates at which incoherent kicks on the rotor occur. Since $\bar{n}_{\rm C}(\bar{n}_{\rm H}+1) < \bar{n}_{\rm H}(\bar{n}_{\rm C}+1)$, the rotor spins up in positive direction. 

The effective master equation \eqref{eq:MEfinal} or \eqref{eq:MErot} is only valid under the assumption of fast thermalization, as compared to the rate at which the rotor rotates. 
To get an idea of the validity regime, we consider steady-state operation of the mill under a weak dissipative load, as introduced in Sec.~\ref{sec:steady}. The approximate model predicts $\la \oL \ra_{\rm ss} = \xi (\bar{n}_{\rm H}-\bar{n}_{\rm C}) /\gamma$ and $\Delta L^2_{\rm ss} = \xi (2\bar{n}_{\rm H}\bar{n}_{\rm C}+\bar{n}_{\rm H}+\bar{n}_{\rm C})/2\gamma + Ik_{\rm B}T$.

The mean angular momentum at steady state is plotted in Fig.~\ref{fig:diffIflywheel} as a function of the thermal bias $\bar{n}_{\rm H}$ at $\bar{n}_{\rm C} = 0$, for different moments of inertia $I$ in the exact model (solid lines) and for the approximate model (dashed). In all cases, the attained momentum grows to a maximum at low bias and then keeps decreasing as the high thermal occupation stalls the energy exchange between the qubits that pushes the rotor. The validity of the approximate model is best at high inertia, i.e.~$\hbar/I \ll \kappa$, and it improves towards higher bias (and poorer work performance).

\section{Qubit versus oscillator piston engine} \label{sec:qubithocompare}

\begin{figure}
   \centering
\centerline{\includegraphics[width=8cm]{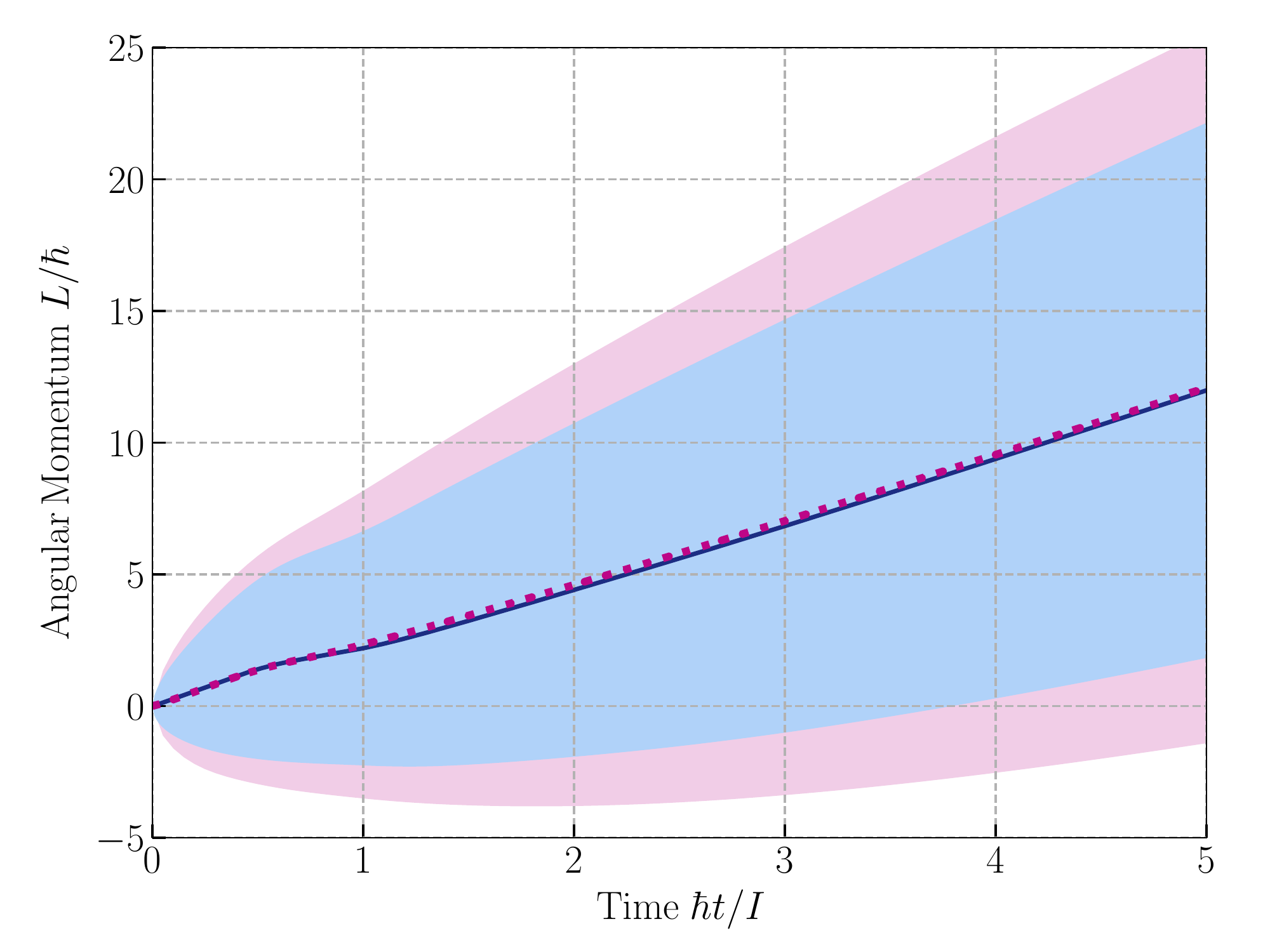}}
\caption{\label{fig:qubitvsho}Comparison of the piston engine dynamics using a qubit (blue) and a harmonic oscillator (red) as working modes. The lines represent the mean angular momenta of the rotor, the shaded regions cover two standard deviations. We chose matching coupling strengths, $g_\mathrm{ho}/\kappa=0.1$ and $g = 3g_\mathrm{ho}$, at $Ig_{ho} = 10\hbar$ and mean bath occupations $(\bar{n}_{\rm H},\bar{n}_{\rm C}) = (1,0)$. Contrary to \cite{alex2017rotor}, the simulations start from the rotor ground state $\ket{\ell =0}$. Angular momenta and harmonic excitations are truncated to $-80 \leq \ell \leq 160$ and $n \leq 7$.} 
\end{figure}

Here, we compare the qubit-driven piston engine model of Sec.~\ref{sec:oldmodel} with the previously studied oscillator-driven case \cite{alex2017rotor}. For this, we replace the spin operators $\hat{\sigma}^+$ and $\hat{\sigma}^-$ with $\hat{a}^\dg$ and $\hat{a}$ respectively. The radiation pressure of a thermally occupied oscillator can accommodate a stronger push than the thermally excited qubit in the interaction term \eqref{eq:HS} at the same coupling strength, since $\mean{\hat{\sigma}^+\hat{\sigma}^-} < \mean{\hat{a}^+ \hat{a}}$. 
Specifically, the instantaneous value \eqref{eq:nphi_qubit}
for the qubit excitation at a given rotor angle $\varphi$ must be compared with the corresponding value for the oscillator excitation number,
\begin{equation}\label{eq:nphi_ho}
n_\mathrm{ho}(\varphi) = \frac{\bar{n}_{\rm H} f_{\rm H}^2\left(\varphi\right)+\bar{n}_{\rm C} f_{\rm C}^2\left(\varphi\right)}{ f_{\rm H}^2\left(\varphi\right)+ f_{\rm C}^2\left(\varphi\right)}.
\end{equation}
Hence for a fair comparison, we here adjust the qubit coupling constant $g$ with respect to the oscillator's $g_{\rm ho}$ so that the maximum push experienced by the piston at $\varphi=\pi/2$ matches, $g = (2\bar{n}_{\rm H} + 1) g_{\rm ho}$. In Fig.~\ref{fig:qubitvsho} we plot the time evolution of the mean angular momentum (lines) and standard deviation (shaded area) for both cases, using the same parameters as in Fig.~\ref{fig:diffworkold}(a). The average behavior is almost the same, with slightly more uncertainty for the oscillator case (red).

\small 
\newcommand{\newblock}{}

\begin{thebibliography}{58}%
\makeatletter
\providecommand \@ifxundefined [1]{%
 \@ifx{#1\undefined}
}%
\providecommand \@ifnum [1]{%
 \ifnum #1\expandafter \@firstoftwo
 \else \expandafter \@secondoftwo
 \fi
}%
\providecommand \@ifx [1]{%
 \ifx #1\expandafter \@firstoftwo
 \else \expandafter \@secondoftwo
 \fi
}%
\providecommand \natexlab [1]{#1}%
\providecommand \enquote  [1]{``#1''}%
\providecommand \bibnamefont  [1]{#1}%
\providecommand \bibfnamefont [1]{#1}%
\providecommand \citenamefont [1]{#1}%
\providecommand \href@noop [0]{\@secondoftwo}%
\providecommand \href [0]{\begingroup \@sanitize@url \@href}%
\providecommand \@href[1]{\@@startlink{#1}\@@href}%
\providecommand \@@href[1]{\endgroup#1\@@endlink}%
\providecommand \@sanitize@url [0]{\catcode `\\12\catcode `\$12\catcode
  `\&12\catcode `\#12\catcode `\^12\catcode `\_12\catcode `\%12\relax}%
\providecommand \@@startlink[1]{}%
\providecommand \@@endlink[0]{}%
\providecommand \url  [0]{\begingroup\@sanitize@url \@url }%
\providecommand \@url [1]{\endgroup\@href {#1}{\urlprefix }}%
\providecommand \urlprefix  [0]{URL }%
\providecommand \Eprint [0]{\href }%
\providecommand \doibase [0]{http://dx.doi.org/}%
\providecommand \selectlanguage [0]{\@gobble}%
\providecommand \bibinfo  [0]{\@secondoftwo}%
\providecommand \bibfield  [0]{\@secondoftwo}%
\providecommand \translation [1]{[#1]}%
\providecommand \BibitemOpen [0]{}%
\providecommand \bibitemStop [0]{}%
\providecommand \bibitemNoStop [0]{.\EOS\space}%
\providecommand \EOS [0]{\spacefactor3000\relax}%
\providecommand \BibitemShut  [1]{\csname bibitem#1\endcsname}%
\let\auto@bib@innerbib\@empty
\bibitem [{\citenamefont {Tonner}\ and\ \citenamefont
  {Mahler}(2005)}]{tonner2005}%
  \BibitemOpen
  \bibfield  {author} {\bibinfo {author} {\bibfnamefont {F.}~\bibnamefont
  {Tonner}}\ and\ \bibinfo {author} {\bibfnamefont {G.}~\bibnamefont
  {Mahler}},\ }\href {\doibase 10.1103/PhysRevE.72.066118} {\bibfield
  {journal} {\bibinfo  {journal} {Phys. Rev. E}\ }\textbf {\bibinfo {volume}
  {72}},\ \bibinfo {pages} {066118} (\bibinfo {year} {2005})}\BibitemShut
  {NoStop}%
\bibitem [{\citenamefont {Youssef}\ \emph {et~al.}(2010)\citenamefont
  {Youssef}, \citenamefont {Mahler},\ and\ \citenamefont
  {Obada}}]{youssef2010}%
  \BibitemOpen
  \bibfield  {author} {\bibinfo {author} {\bibfnamefont {M.}~\bibnamefont
  {Youssef}}, \bibinfo {author} {\bibfnamefont {G.}~\bibnamefont {Mahler}}, \
  and\ \bibinfo {author} {\bibfnamefont {A.-S.}\ \bibnamefont {Obada}},\ }\href
  {\doibase https://doi.org/10.1016/j.physe.2009.06.032} {\bibfield  {journal}
  {\bibinfo  {journal} {Physica E}\ }\textbf {\bibinfo {volume} {42}},\
  \bibinfo {pages} {454 } (\bibinfo {year} {2010})}\BibitemShut {NoStop}%
\bibitem [{\citenamefont {Brunner}\ \emph {et~al.}(2012)\citenamefont
  {Brunner}, \citenamefont {Linden}, \citenamefont {Popescu},\ and\
  \citenamefont {Skrzypczyk}}]{brunner2012virtual}%
  \BibitemOpen
  \bibfield  {author} {\bibinfo {author} {\bibfnamefont {N.}~\bibnamefont
  {Brunner}}, \bibinfo {author} {\bibfnamefont {N.}~\bibnamefont {Linden}},
  \bibinfo {author} {\bibfnamefont {S.}~\bibnamefont {Popescu}}, \ and\
  \bibinfo {author} {\bibfnamefont {P.}~\bibnamefont {Skrzypczyk}},\ }\href
  {\doibase 10.1103/PhysRevE.85.051117} {\bibfield  {journal} {\bibinfo
  {journal} {Phys. Rev. E}\ }\textbf {\bibinfo {volume} {85}},\ \bibinfo
  {pages} {051117} (\bibinfo {year} {2012})}\BibitemShut {NoStop}%
\bibitem [{\citenamefont {Gilz}\ \emph {et~al.}(2013)\citenamefont {Gilz},
  \citenamefont {Thesing},\ and\ \citenamefont {Anglin}}]{gilz2013}%
  \BibitemOpen
  \bibfield  {author} {\bibinfo {author} {\bibfnamefont {L.}~\bibnamefont
  {Gilz}}, \bibinfo {author} {\bibfnamefont {E.~P.}\ \bibnamefont {Thesing}}, \
  and\ \bibinfo {author} {\bibfnamefont {J.~R.}\ \bibnamefont {Anglin}},\
  }\href {https://arxiv.org/abs/1304.3222} {\bibfield  {journal} {\bibinfo
  {journal} {arXiv:1304.3222}\ } (\bibinfo {year} {2013})}\BibitemShut
  {NoStop}%
\bibitem [{\citenamefont {Mari}\ \emph {et~al.}(2015)\citenamefont {Mari},
  \citenamefont {Farace},\ and\ \citenamefont {Giovannetti}}]{mari2015quantum}%
  \BibitemOpen
  \bibfield  {author} {\bibinfo {author} {\bibfnamefont {A.}~\bibnamefont
  {Mari}}, \bibinfo {author} {\bibfnamefont {A.}~\bibnamefont {Farace}}, \ and\
  \bibinfo {author} {\bibfnamefont {V.}~\bibnamefont {Giovannetti}},\ }\href
  {http://stacks.iop.org/0953-4075/48/i=17/a=175501} {\bibfield  {journal}
  {\bibinfo  {journal} {J. Phys. B}\ }\textbf {\bibinfo {volume} {48}},\
  \bibinfo {pages} {175501} (\bibinfo {year} {2015})}\BibitemShut {NoStop}%
\bibitem [{\citenamefont {Levy}\ \emph {et~al.}(2016)\citenamefont {Levy},
  \citenamefont {Di\'osi},\ and\ \citenamefont
  {Kosloff}}]{kosloff2016flywheel}%
  \BibitemOpen
  \bibfield  {author} {\bibinfo {author} {\bibfnamefont {A.}~\bibnamefont
  {Levy}}, \bibinfo {author} {\bibfnamefont {L.}~\bibnamefont {Di\'osi}}, \
  and\ \bibinfo {author} {\bibfnamefont {R.}~\bibnamefont {Kosloff}},\ }\href
  {\doibase 10.1103/PhysRevA.93.052119} {\bibfield  {journal} {\bibinfo
  {journal} {Phys. Rev. A}\ }\textbf {\bibinfo {volume} {93}},\ \bibinfo
  {pages} {052119} (\bibinfo {year} {2016})}\BibitemShut {NoStop}%
\bibitem [{\citenamefont {Roulet}\ \emph {et~al.}(2017)\citenamefont {Roulet},
  \citenamefont {Nimmrichter}, \citenamefont {Arrazola}, \citenamefont {Seah},\
  and\ \citenamefont {Scarani}}]{alex2017rotor}%
  \BibitemOpen
  \bibfield  {author} {\bibinfo {author} {\bibfnamefont {A.}~\bibnamefont
  {Roulet}}, \bibinfo {author} {\bibfnamefont {S.}~\bibnamefont {Nimmrichter}},
  \bibinfo {author} {\bibfnamefont {J.~M.}\ \bibnamefont {Arrazola}}, \bibinfo
  {author} {\bibfnamefont {S.}~\bibnamefont {Seah}}, \ and\ \bibinfo {author}
  {\bibfnamefont {V.}~\bibnamefont {Scarani}},\ }\href {\doibase
  10.1103/PhysRevE.95.062131} {\bibfield  {journal} {\bibinfo  {journal} {Phys.
  Rev. E}\ }\textbf {\bibinfo {volume} {95}},\ \bibinfo {pages} {062131}
  (\bibinfo {year} {2017})}\BibitemShut {NoStop}%
\bibitem [{\citenamefont {Hardal}\ \emph {et~al.}(2017)\citenamefont {Hardal},
  \citenamefont {Aslan}, \citenamefont {Wilson},\ and\ \citenamefont
  {M\"ustecapl\ifmmode \imath \else \i \fi{}o\ifmmode~\breve{g}\else
  \u{g}\fi{}lu}}]{hardal2017}%
  \BibitemOpen
  \bibfield  {author} {\bibinfo {author} {\bibfnamefont {A.~U.~C.}\
  \bibnamefont {Hardal}}, \bibinfo {author} {\bibfnamefont {N.}~\bibnamefont
  {Aslan}}, \bibinfo {author} {\bibfnamefont {C.~M.}\ \bibnamefont {Wilson}}, \
  and\ \bibinfo {author} {\bibfnamefont {O.~E.}\ \bibnamefont
  {M\"ustecapl\ifmmode \imath \else \i \fi{}o\ifmmode~\breve{g}\else
  \u{g}\fi{}lu}},\ }\href {\doibase 10.1103/PhysRevE.96.062120} {\bibfield
  {journal} {\bibinfo  {journal} {Phys. Rev. E}\ }\textbf {\bibinfo {volume}
  {96}},\ \bibinfo {pages} {062120} (\bibinfo {year} {2017})}\BibitemShut
  {NoStop}%
\bibitem [{\citenamefont {Kosloff}(1984)}]{kosloff1984}%
  \BibitemOpen
  \bibfield  {author} {\bibinfo {author} {\bibfnamefont {R.}~\bibnamefont
  {Kosloff}},\ }\href {\doibase 10.1063/1.446862} {\bibfield  {journal}
  {\bibinfo  {journal} {J. Chem. Phys.}\ }\textbf {\bibinfo {volume} {80}},\
  \bibinfo {pages} {1625} (\bibinfo {year} {1984})}\BibitemShut {NoStop}%
\bibitem [{\citenamefont {Scully}\ \emph {et~al.}(2003)\citenamefont {Scully},
  \citenamefont {Zubairy}, \citenamefont {Agarwal},\ and\ \citenamefont
  {Walther}}]{scully2003}%
  \BibitemOpen
  \bibfield  {author} {\bibinfo {author} {\bibfnamefont {M.~O.}\ \bibnamefont
  {Scully}}, \bibinfo {author} {\bibfnamefont {M.~S.}\ \bibnamefont {Zubairy}},
  \bibinfo {author} {\bibfnamefont {G.~S.}\ \bibnamefont {Agarwal}}, \ and\
  \bibinfo {author} {\bibfnamefont {H.}~\bibnamefont {Walther}},\ }\href
  {\doibase 10.1126/science.1078955} {\bibfield  {journal} {\bibinfo  {journal}
  {Science}\ }\textbf {\bibinfo {volume} {299}},\ \bibinfo {pages} {862}
  (\bibinfo {year} {2003})}\BibitemShut {NoStop}%
\bibitem [{\citenamefont {Rezek}\ and\ \citenamefont
  {Kosloff}(2006)}]{rezek2006}%
  \BibitemOpen
  \bibfield  {author} {\bibinfo {author} {\bibfnamefont {Y.}~\bibnamefont
  {Rezek}}\ and\ \bibinfo {author} {\bibfnamefont {R.}~\bibnamefont
  {Kosloff}},\ }\href {http://stacks.iop.org/1367-2630/8/i=5/a=083} {\bibfield
  {journal} {\bibinfo  {journal} {New J. Phys.}\ }\textbf {\bibinfo {volume}
  {8}},\ \bibinfo {pages} {83} (\bibinfo {year} {2006})}\BibitemShut {NoStop}%
\bibitem [{\citenamefont {Alicki}(2014)}]{alicki2014}%
  \BibitemOpen
  \bibfield  {author} {\bibinfo {author} {\bibfnamefont {R.}~\bibnamefont
  {Alicki}},\ }\href {\doibase 10.1142/S1230161214400022} {\bibfield  {journal}
  {\bibinfo  {journal} {Open Syst. Inf. Dyn.}\ }\textbf {\bibinfo {volume}
  {21}},\ \bibinfo {pages} {1440002} (\bibinfo {year} {2014})}\BibitemShut
  {NoStop}%
\bibitem [{\citenamefont {Zhang}\ \emph {et~al.}(2014)\citenamefont {Zhang},
  \citenamefont {Bariani},\ and\ \citenamefont {Meystre}}]{zhang2014}%
  \BibitemOpen
  \bibfield  {author} {\bibinfo {author} {\bibfnamefont {K.}~\bibnamefont
  {Zhang}}, \bibinfo {author} {\bibfnamefont {F.}~\bibnamefont {Bariani}}, \
  and\ \bibinfo {author} {\bibfnamefont {P.}~\bibnamefont {Meystre}},\ }\href
  {\doibase 10.1103/PhysRevLett.112.150602} {\bibfield  {journal} {\bibinfo
  {journal} {Phys. Rev. Lett.}\ }\textbf {\bibinfo {volume} {112}},\ \bibinfo
  {pages} {150602} (\bibinfo {year} {2014})}\BibitemShut {NoStop}%
\bibitem [{\citenamefont {Uzdin}\ \emph {et~al.}(2016)\citenamefont {Uzdin},
  \citenamefont {Levy},\ and\ \citenamefont {Kosloff}}]{uzdin2016}%
  \BibitemOpen
  \bibfield  {author} {\bibinfo {author} {\bibfnamefont {R.}~\bibnamefont
  {Uzdin}}, \bibinfo {author} {\bibfnamefont {A.}~\bibnamefont {Levy}}, \ and\
  \bibinfo {author} {\bibfnamefont {R.}~\bibnamefont {Kosloff}},\ }\href
  {\doibase 10.3390/e18040124} {\bibfield  {journal} {\bibinfo  {journal}
  {Entropy}\ }\textbf {\bibinfo {volume} {18}} (\bibinfo {year} {2016}),\
  10.3390/e18040124}\BibitemShut {NoStop}%
\bibitem [{\citenamefont {Roncaglia}\ \emph {et~al.}(2014)\citenamefont
  {Roncaglia}, \citenamefont {Cerisola},\ and\ \citenamefont
  {Paz}}]{roncaglia2014}%
  \BibitemOpen
  \bibfield  {author} {\bibinfo {author} {\bibfnamefont {A.~J.}\ \bibnamefont
  {Roncaglia}}, \bibinfo {author} {\bibfnamefont {F.}~\bibnamefont {Cerisola}},
  \ and\ \bibinfo {author} {\bibfnamefont {J.~P.}\ \bibnamefont {Paz}},\ }\href
  {\doibase 10.1103/PhysRevLett.113.250601} {\bibfield  {journal} {\bibinfo
  {journal} {Phys. Rev. Lett.}\ }\textbf {\bibinfo {volume} {113}},\ \bibinfo
  {pages} {250601} (\bibinfo {year} {2014})}\BibitemShut {NoStop}%
\bibitem [{\citenamefont {Talkner}\ and\ \citenamefont
  {H\"anggi}(2016)}]{hanggi2016}%
  \BibitemOpen
  \bibfield  {author} {\bibinfo {author} {\bibfnamefont {P.}~\bibnamefont
  {Talkner}}\ and\ \bibinfo {author} {\bibfnamefont {P.}~\bibnamefont
  {H\"anggi}},\ }\href {\doibase 10.1103/PhysRevE.93.022131} {\bibfield
  {journal} {\bibinfo  {journal} {Phys. Rev. E}\ }\textbf {\bibinfo {volume}
  {93}},\ \bibinfo {pages} {022131} (\bibinfo {year} {2016})}\BibitemShut
  {NoStop}%
\bibitem [{\citenamefont {Vinjanampathy}\ and\ \citenamefont
  {Anders}(2016)}]{sai2016review}%
  \BibitemOpen
  \bibfield  {author} {\bibinfo {author} {\bibfnamefont {S.}~\bibnamefont
  {Vinjanampathy}}\ and\ \bibinfo {author} {\bibfnamefont {J.}~\bibnamefont
  {Anders}},\ }\href {\doibase 10.1080/00107514.2016.1201896} {\bibfield
  {journal} {\bibinfo  {journal} {Contemp. Phys.}\ }\textbf {\bibinfo {volume}
  {57}},\ \bibinfo {pages} {545} (\bibinfo {year} {2016})}\BibitemShut
  {NoStop}%
\bibitem [{\citenamefont {Hayashi}\ and\ \citenamefont
  {Tajima}(2017)}]{masahito2017}%
  \BibitemOpen
  \bibfield  {author} {\bibinfo {author} {\bibfnamefont {M.}~\bibnamefont
  {Hayashi}}\ and\ \bibinfo {author} {\bibfnamefont {H.}~\bibnamefont
  {Tajima}},\ }\href {\doibase 10.1103/PhysRevA.95.032132} {\bibfield
  {journal} {\bibinfo  {journal} {Phys. Rev. A}\ }\textbf {\bibinfo {volume}
  {95}},\ \bibinfo {pages} {032132} (\bibinfo {year} {2017})}\BibitemShut
  {NoStop}%
\bibitem [{\citenamefont {Talkner}\ \emph {et~al.}(2007)\citenamefont
  {Talkner}, \citenamefont {Lutz},\ and\ \citenamefont
  {H\"anggi}}]{hanggi2007}%
  \BibitemOpen
  \bibfield  {author} {\bibinfo {author} {\bibfnamefont {P.}~\bibnamefont
  {Talkner}}, \bibinfo {author} {\bibfnamefont {E.}~\bibnamefont {Lutz}}, \
  and\ \bibinfo {author} {\bibfnamefont {P.}~\bibnamefont {H\"anggi}},\ }\href
  {\doibase 10.1103/PhysRevE.75.050102} {\bibfield  {journal} {\bibinfo
  {journal} {Phys. Rev. E}\ }\textbf {\bibinfo {volume} {75}},\ \bibinfo
  {pages} {050102} (\bibinfo {year} {2007})}\BibitemShut {NoStop}%
\bibitem [{\citenamefont {Allahverdyan}\ \emph {et~al.}(2004)\citenamefont
  {Allahverdyan}, \citenamefont {Balian},\ and\ \citenamefont
  {Nieuwenhuizen}}]{allah2004work}%
  \BibitemOpen
  \bibfield  {author} {\bibinfo {author} {\bibfnamefont {A.~E.}\ \bibnamefont
  {Allahverdyan}}, \bibinfo {author} {\bibfnamefont {R.}~\bibnamefont
  {Balian}}, \ and\ \bibinfo {author} {\bibfnamefont {T.~M.}\ \bibnamefont
  {Nieuwenhuizen}},\ }\href {http://stacks.iop.org/0295-5075/67/i=4/a=565}
  {\bibfield  {journal} {\bibinfo  {journal} {Europhys. Lett.}\ }\textbf
  {\bibinfo {volume} {67}},\ \bibinfo {pages} {565} (\bibinfo {year}
  {2004})}\BibitemShut {NoStop}%
\bibitem [{\citenamefont {Skrzypczyk}\ \emph {et~al.}(2014)\citenamefont
  {Skrzypczyk}, \citenamefont {Short},\ and\ \citenamefont
  {Popescu}}]{skrzypczyk2014work}%
  \BibitemOpen
  \bibfield  {author} {\bibinfo {author} {\bibfnamefont {P.}~\bibnamefont
  {Skrzypczyk}}, \bibinfo {author} {\bibfnamefont {A.~J.}\ \bibnamefont
  {Short}}, \ and\ \bibinfo {author} {\bibfnamefont {S.}~\bibnamefont
  {Popescu}},\ }\href {http://dx.doi.org/10.1038/ncomms5185} {\bibfield
  {journal} {\bibinfo  {journal} {Nat. Commun.}\ }\textbf {\bibinfo {volume}
  {5}},\ \bibinfo {pages} {4185} (\bibinfo {year} {2014})}\BibitemShut
  {NoStop}%
\bibitem [{\citenamefont {Binder}\ \emph {et~al.}(2015)\citenamefont {Binder},
  \citenamefont {Vinjanampathy}, \citenamefont {Modi},\ and\ \citenamefont
  {Goold}}]{binder2015work}%
  \BibitemOpen
  \bibfield  {author} {\bibinfo {author} {\bibfnamefont {F.}~\bibnamefont
  {Binder}}, \bibinfo {author} {\bibfnamefont {S.}~\bibnamefont
  {Vinjanampathy}}, \bibinfo {author} {\bibfnamefont {K.}~\bibnamefont {Modi}},
  \ and\ \bibinfo {author} {\bibfnamefont {J.}~\bibnamefont {Goold}},\ }\href
  {\doibase 10.1103/PhysRevE.91.032119} {\bibfield  {journal} {\bibinfo
  {journal} {Phys. Rev. E}\ }\textbf {\bibinfo {volume} {91}},\ \bibinfo
  {pages} {032119} (\bibinfo {year} {2015})}\BibitemShut {NoStop}%
\bibitem [{\citenamefont {Perarnau-Llobet}\ \emph {et~al.}(2015)\citenamefont
  {Perarnau-Llobet}, \citenamefont {Hovhannisyan}, \citenamefont {Huber},
  \citenamefont {Skrzypczyk}, \citenamefont {Brunner},\ and\ \citenamefont
  {Ac\'{\i}n}}]{acin2015}%
  \BibitemOpen
  \bibfield  {author} {\bibinfo {author} {\bibfnamefont {M.}~\bibnamefont
  {Perarnau-Llobet}}, \bibinfo {author} {\bibfnamefont {K.~V.}\ \bibnamefont
  {Hovhannisyan}}, \bibinfo {author} {\bibfnamefont {M.}~\bibnamefont {Huber}},
  \bibinfo {author} {\bibfnamefont {P.}~\bibnamefont {Skrzypczyk}}, \bibinfo
  {author} {\bibfnamefont {N.}~\bibnamefont {Brunner}}, \ and\ \bibinfo
  {author} {\bibfnamefont {A.}~\bibnamefont {Ac\'{\i}n}},\ }\href {\doibase
  10.1103/PhysRevX.5.041011} {\bibfield  {journal} {\bibinfo  {journal} {Phys.
  Rev. X}\ }\textbf {\bibinfo {volume} {5}},\ \bibinfo {pages} {041011}
  (\bibinfo {year} {2015})}\BibitemShut {NoStop}%
\bibitem [{\citenamefont {Niedenzu}\ \emph {et~al.}(2018)\citenamefont
  {Niedenzu}, \citenamefont {Mukherjee}, \citenamefont {Ghosh}, \citenamefont
  {Kofman},\ and\ \citenamefont {Kurizki}}]{niedenzu2017work}%
  \BibitemOpen
  \bibfield  {author} {\bibinfo {author} {\bibfnamefont {W.}~\bibnamefont
  {Niedenzu}}, \bibinfo {author} {\bibfnamefont {V.}~\bibnamefont {Mukherjee}},
  \bibinfo {author} {\bibfnamefont {A.}~\bibnamefont {Ghosh}}, \bibinfo
  {author} {\bibfnamefont {A.~G.}\ \bibnamefont {Kofman}}, \ and\ \bibinfo
  {author} {\bibfnamefont {G.}~\bibnamefont {Kurizki}},\ }\href {\doibase
  10.1038/s41467-017-01991-6} {\bibfield  {journal} {\bibinfo  {journal} {Nat.
  Commun.}\ }\textbf {\bibinfo {volume} {9}},\ \bibinfo {pages} {165} (\bibinfo
  {year} {2018})}\BibitemShut {NoStop}%
\bibitem [{\citenamefont {Linden}\ \emph {et~al.}(2010)\citenamefont {Linden},
  \citenamefont {Popescu},\ and\ \citenamefont
  {Skrzypczyk}}]{linden2010fridge}%
  \BibitemOpen
  \bibfield  {author} {\bibinfo {author} {\bibfnamefont {N.}~\bibnamefont
  {Linden}}, \bibinfo {author} {\bibfnamefont {S.}~\bibnamefont {Popescu}}, \
  and\ \bibinfo {author} {\bibfnamefont {P.}~\bibnamefont {Skrzypczyk}},\
  }\href {\doibase 10.1103/PhysRevLett.105.130401} {\bibfield  {journal}
  {\bibinfo  {journal} {Phys. Rev. Lett.}\ }\textbf {\bibinfo {volume} {105}},\
  \bibinfo {pages} {130401} (\bibinfo {year} {2010})}\BibitemShut {NoStop}%
\bibitem [{\citenamefont {Levy}\ and\ \citenamefont
  {Kosloff}(2012{\natexlab{a}})}]{levy2012fridge}%
  \BibitemOpen
  \bibfield  {author} {\bibinfo {author} {\bibfnamefont {A.}~\bibnamefont
  {Levy}}\ and\ \bibinfo {author} {\bibfnamefont {R.}~\bibnamefont {Kosloff}},\
  }\href {\doibase 10.1103/PhysRevLett.108.070604} {\bibfield  {journal}
  {\bibinfo  {journal} {Phys. Rev. Lett.}\ }\textbf {\bibinfo {volume} {108}},\
  \bibinfo {pages} {070604} (\bibinfo {year} {2012}{\natexlab{a}})}\BibitemShut
  {NoStop}%
\bibitem [{\citenamefont {Brask}\ and\ \citenamefont
  {Brunner}(2015)}]{brask2015}%
  \BibitemOpen
  \bibfield  {author} {\bibinfo {author} {\bibfnamefont {J.~B.}\ \bibnamefont
  {Brask}}\ and\ \bibinfo {author} {\bibfnamefont {N.}~\bibnamefont
  {Brunner}},\ }\href {\doibase 10.1103/PhysRevE.92.062101} {\bibfield
  {journal} {\bibinfo  {journal} {Phys. Rev. E}\ }\textbf {\bibinfo {volume}
  {92}},\ \bibinfo {pages} {062101} (\bibinfo {year} {2015})}\BibitemShut
  {NoStop}%
\bibitem [{\citenamefont {Mitchison}\ \emph {et~al.}(2015)\citenamefont
  {Mitchison}, \citenamefont {Woods}, \citenamefont {Prior},\ and\
  \citenamefont {Huber}}]{mitchisonFridge2015}%
  \BibitemOpen
  \bibfield  {author} {\bibinfo {author} {\bibfnamefont {M.~T.}\ \bibnamefont
  {Mitchison}}, \bibinfo {author} {\bibfnamefont {M.~P.}\ \bibnamefont
  {Woods}}, \bibinfo {author} {\bibfnamefont {J.}~\bibnamefont {Prior}}, \ and\
  \bibinfo {author} {\bibfnamefont {M.}~\bibnamefont {Huber}},\ }\href
  {\doibase 10.1088/1367-2630/17/11/115013} {\bibfield  {journal} {\bibinfo
  {journal} {New J. Phys.}\ }\textbf {\bibinfo {volume} {17}},\ \bibinfo
  {pages} {115013} (\bibinfo {year} {2015})}\BibitemShut {NoStop}%
\bibitem [{\citenamefont {Stickler}\ \emph {et~al.}(2017)\citenamefont
  {Stickler}, \citenamefont {Schrinski},\ and\ \citenamefont
  {Hornberger}}]{benjamin2017}%
  \BibitemOpen
  \bibfield  {author} {\bibinfo {author} {\bibfnamefont {B.~A.}\ \bibnamefont
  {Stickler}}, \bibinfo {author} {\bibfnamefont {B.}~\bibnamefont {Schrinski}},
  \ and\ \bibinfo {author} {\bibfnamefont {K.}~\bibnamefont {Hornberger}},\
  }\href {https://arxiv.org/abs/1712.05163} {\bibfield  {journal} {\bibinfo
  {journal} {arXiv:1712.05163}\ } (\bibinfo {year} {2017})}\BibitemShut
  {NoStop}%
\bibitem [{\citenamefont {Alicki}(1979)}]{alicki1979work}%
  \BibitemOpen
  \bibfield  {author} {\bibinfo {author} {\bibfnamefont {R.}~\bibnamefont
  {Alicki}},\ }\href {http://stacks.iop.org/0305-4470/12/i=5/a=007} {\bibfield
  {journal} {\bibinfo  {journal} {J. Phys. A}\ }\textbf {\bibinfo {volume}
  {12}},\ \bibinfo {pages} {L103} (\bibinfo {year} {1979})}\BibitemShut
  {NoStop}%
\bibitem [{\citenamefont {Goold}\ \emph {et~al.}(2016)\citenamefont {Goold},
  \citenamefont {Huber}, \citenamefont {Riera}, \citenamefont {del Rio},\ and\
  \citenamefont {Skrzypczyk}}]{goold2016review}%
  \BibitemOpen
  \bibfield  {author} {\bibinfo {author} {\bibfnamefont {J.}~\bibnamefont
  {Goold}}, \bibinfo {author} {\bibfnamefont {M.}~\bibnamefont {Huber}},
  \bibinfo {author} {\bibfnamefont {A.}~\bibnamefont {Riera}}, \bibinfo
  {author} {\bibfnamefont {L.}~\bibnamefont {del Rio}}, \ and\ \bibinfo
  {author} {\bibfnamefont {P.}~\bibnamefont {Skrzypczyk}},\ }\href
  {http://stacks.iop.org/1751-8121/49/i=14/a=143001} {\bibfield  {journal}
  {\bibinfo  {journal} {J. Phys. A}\ }\textbf {\bibinfo {volume} {49}},\
  \bibinfo {pages} {143001} (\bibinfo {year} {2016})}\BibitemShut {NoStop}%
\bibitem [{\citenamefont {Lenard}(1978)}]{lenard1978work}%
  \BibitemOpen
  \bibfield  {author} {\bibinfo {author} {\bibfnamefont {A.}~\bibnamefont
  {Lenard}},\ }\href {\doibase 10.1007/BF01011769} {\bibfield  {journal}
  {\bibinfo  {journal} {J. Stat. Phys.}\ }\textbf {\bibinfo {volume} {19}},\
  \bibinfo {pages} {575} (\bibinfo {year} {1978})}\BibitemShut {NoStop}%
\bibitem [{\citenamefont {Pusz}\ and\ \citenamefont
  {Woronowicz}(1978)}]{pusz1978passive}%
  \BibitemOpen
  \bibfield  {author} {\bibinfo {author} {\bibfnamefont {W.}~\bibnamefont
  {Pusz}}\ and\ \bibinfo {author} {\bibfnamefont {S.~L.}\ \bibnamefont
  {Woronowicz}},\ }\href {\doibase 10.1007/BF01614224} {\bibfield  {journal}
  {\bibinfo  {journal} {Commun. Math. Phys.}\ }\textbf {\bibinfo {volume}
  {58}},\ \bibinfo {pages} {273} (\bibinfo {year} {1978})}\BibitemShut
  {NoStop}%
\bibitem [{\citenamefont {Johansson}\ \emph {et~al.}(2013)\citenamefont
  {Johansson}, \citenamefont {Nation},\ and\ \citenamefont
  {Nori}}]{johansson2013qutip}%
  \BibitemOpen
  \bibfield  {author} {\bibinfo {author} {\bibfnamefont {J.}~\bibnamefont
  {Johansson}}, \bibinfo {author} {\bibfnamefont {P.}~\bibnamefont {Nation}}, \
  and\ \bibinfo {author} {\bibfnamefont {F.}~\bibnamefont {Nori}},\ }\href
  {\doibase 10.1016/j.cpc.2012.11.019} {\bibfield  {journal} {\bibinfo
  {journal} {Comput. Phys. Commun.}\ }\textbf {\bibinfo {volume} {184}},\
  \bibinfo {pages} {1234} (\bibinfo {year} {2013})}\BibitemShut {NoStop}%
\bibitem [{\citenamefont {Chartrand}(2011)}]{chartrand2011noisydiff}%
  \BibitemOpen
  \bibfield  {author} {\bibinfo {author} {\bibfnamefont {R.}~\bibnamefont
  {Chartrand}},\ }\href {\doibase 10.5402/2011/164564} {\bibfield  {journal}
  {\bibinfo  {journal} {ISRN Appl. Math.}\ }\textbf {\bibinfo {volume} {2011}}
  (\bibinfo {year} {2011}),\ 10.5402/2011/164564}\BibitemShut {NoStop}%
\bibitem [{\citenamefont {Gardiner}\ and\ \citenamefont
  {Collett}(1985)}]{gardiner1985}%
  \BibitemOpen
  \bibfield  {author} {\bibinfo {author} {\bibfnamefont {C.~W.}\ \bibnamefont
  {Gardiner}}\ and\ \bibinfo {author} {\bibfnamefont {M.~J.}\ \bibnamefont
  {Collett}},\ }\href {\doibase 10.1103/PhysRevA.31.3761} {\bibfield  {journal}
  {\bibinfo  {journal} {Phys. Rev. A}\ }\textbf {\bibinfo {volume} {31}},\
  \bibinfo {pages} {3761} (\bibinfo {year} {1985})}\BibitemShut {NoStop}%
\bibitem [{\citenamefont {Breuer}\ and\ \citenamefont
  {Petruccione}(2002)}]{Breuer2002}%
  \BibitemOpen
  \bibfield  {author} {\bibinfo {author} {\bibfnamefont {H.-P.}\ \bibnamefont
  {Breuer}}\ and\ \bibinfo {author} {\bibfnamefont {F.}~\bibnamefont
  {Petruccione}},\ }\href
  {http://books.google.com/books?id=0Yx5VzaMYm8C{\&}pgis=1} {\emph {\bibinfo
  {title} {{The Theory of Open Quantum Systems}}}}\ (\bibinfo  {publisher}
  {Oxford University Press},\ \bibinfo {year} {2002})\BibitemShut {NoStop}%
\bibitem [{\citenamefont {Carmichael}(2003)}]{carmichael2003}%
  \BibitemOpen
  \bibfield  {author} {\bibinfo {author} {\bibfnamefont {H.~J.}\ \bibnamefont
  {Carmichael}},\ }\href {http://www.springer.com/gp/book/9783540548829} {\emph
  {\bibinfo {title} {Statistical Methods in Quantum Optics 1: Master Equations
  and Fokker-Planck Equations}}}\ (\bibinfo  {publisher} {Springer},\ \bibinfo
  {year} {2003})\BibitemShut {NoStop}%
\bibitem [{\citenamefont {Gardiner}\ and\ \citenamefont
  {Zoller}(2004)}]{gardiner2004noise}%
  \BibitemOpen
  \bibfield  {author} {\bibinfo {author} {\bibfnamefont {C.}~\bibnamefont
  {Gardiner}}\ and\ \bibinfo {author} {\bibfnamefont {P.}~\bibnamefont
  {Zoller}},\ }\href {http://www.springer.com/gp/book/9783540223016} {\emph
  {\bibinfo {title} {Quantum Noise}}}\ (\bibinfo  {publisher} {Springer},\
  \bibinfo {year} {2004})\BibitemShut {NoStop}%
\bibitem [{\citenamefont {Rivas}\ \emph {et~al.}(2010)\citenamefont {Rivas},
  \citenamefont {Plato}, \citenamefont {Huelga},\ and\ \citenamefont
  {Plenio}}]{rivas2010}%
  \BibitemOpen
  \bibfield  {author} {\bibinfo {author} {\bibfnamefont {A.}~\bibnamefont
  {Rivas}}, \bibinfo {author} {\bibfnamefont {A.~D.~K.}\ \bibnamefont {Plato}},
  \bibinfo {author} {\bibfnamefont {S.~F.}\ \bibnamefont {Huelga}}, \ and\
  \bibinfo {author} {\bibfnamefont {M.~B.}\ \bibnamefont {Plenio}},\ }\href
  {\doibase 10.1088/1367-2630/12/11/113032} {\bibfield  {journal} {\bibinfo
  {journal} {New J. Phys.}\ }\textbf {\bibinfo {volume} {12}},\ \bibinfo
  {pages} {113032} (\bibinfo {year} {2010})}\BibitemShut {NoStop}%
\bibitem [{\citenamefont {Levy}\ and\ \citenamefont
  {Kosloff}(2014)}]{levy2014}%
  \BibitemOpen
  \bibfield  {author} {\bibinfo {author} {\bibfnamefont {A.}~\bibnamefont
  {Levy}}\ and\ \bibinfo {author} {\bibfnamefont {R.}~\bibnamefont {Kosloff}},\
  }\href {\doibase 10.1209/0295-5075/107/20004} {\bibfield  {journal} {\bibinfo
   {journal} {Europhys. Lett.}\ }\textbf {\bibinfo {volume} {107}},\ \bibinfo
  {pages} {20004} (\bibinfo {year} {2014})}\BibitemShut {NoStop}%
\bibitem [{\citenamefont {Hofer}\ \emph {et~al.}(2017)\citenamefont {Hofer},
  \citenamefont {Perarnau-Llobet}, \citenamefont {Miranda}, \citenamefont
  {Haack}, \citenamefont {Silva}, \citenamefont {Brask},\ and\ \citenamefont
  {Brunner}}]{hofer2017}%
  \BibitemOpen
  \bibfield  {author} {\bibinfo {author} {\bibfnamefont {P.~P.}\ \bibnamefont
  {Hofer}}, \bibinfo {author} {\bibfnamefont {M.}~\bibnamefont
  {Perarnau-Llobet}}, \bibinfo {author} {\bibfnamefont {L.~D.~M.}\ \bibnamefont
  {Miranda}}, \bibinfo {author} {\bibfnamefont {G.}~\bibnamefont {Haack}},
  \bibinfo {author} {\bibfnamefont {R.}~\bibnamefont {Silva}}, \bibinfo
  {author} {\bibfnamefont {J.~B.}\ \bibnamefont {Brask}}, \ and\ \bibinfo
  {author} {\bibfnamefont {N.}~\bibnamefont {Brunner}},\ }\href {\doibase
  10.1088/1367-2630/aa964f} {\bibfield  {journal} {\bibinfo  {journal} {New J.
  Phys.}\ }\textbf {\bibinfo {volume} {19}},\ \bibinfo {pages} {123037}
  (\bibinfo {year} {2017})}\BibitemShut {NoStop}%
\bibitem [{\citenamefont {Gonz{\'{a}}lez}\ \emph {et~al.}(2017)\citenamefont
  {Gonz{\'{a}}lez}, \citenamefont {Correa}, \citenamefont {Nocerino},
  \citenamefont {Palao}, \citenamefont {Alonso},\ and\ \citenamefont
  {Adesso}}]{gonzalez2017}%
  \BibitemOpen
  \bibfield  {author} {\bibinfo {author} {\bibfnamefont {J.~O.}\ \bibnamefont
  {Gonz{\'{a}}lez}}, \bibinfo {author} {\bibfnamefont {L.~A.}\ \bibnamefont
  {Correa}}, \bibinfo {author} {\bibfnamefont {G.}~\bibnamefont {Nocerino}},
  \bibinfo {author} {\bibfnamefont {J.~P.}\ \bibnamefont {Palao}}, \bibinfo
  {author} {\bibfnamefont {D.}~\bibnamefont {Alonso}}, \ and\ \bibinfo {author}
  {\bibfnamefont {G.}~\bibnamefont {Adesso}},\ }\href {\doibase
  10.1142/S1230161217400108} {\bibfield  {journal} {\bibinfo  {journal} {Open
  Syst. Inf. Dyn.}\ }\textbf {\bibinfo {volume} {24}},\ \bibinfo {pages}
  {1740010} (\bibinfo {year} {2017})}\BibitemShut {NoStop}%
\bibitem [{\citenamefont {Chen}\ \emph {et~al.}(1997)\citenamefont {Chen},
  \citenamefont {Sun},\ and\ \citenamefont {Wu}}]{chen1997}%
  \BibitemOpen
  \bibfield  {author} {\bibinfo {author} {\bibfnamefont {L.}~\bibnamefont
  {Chen}}, \bibinfo {author} {\bibfnamefont {F.}~\bibnamefont {Sun}}, \ and\
  \bibinfo {author} {\bibfnamefont {C.}~\bibnamefont {Wu}},\ }\href {\doibase
  https://doi.org/10.1016/S0196-8904(96)00103-3} {\bibfield  {journal}
  {\bibinfo  {journal} {Energy Convers. Manag.}\ }\textbf {\bibinfo {volume}
  {38}},\ \bibinfo {pages} {1501 } (\bibinfo {year} {1997})}\BibitemShut
  {NoStop}%
\bibitem [{\citenamefont {Levy}\ and\ \citenamefont
  {Kosloff}(2012{\natexlab{b}})}]{levy2012quantum}%
  \BibitemOpen
  \bibfield  {author} {\bibinfo {author} {\bibfnamefont {A.}~\bibnamefont
  {Levy}}\ and\ \bibinfo {author} {\bibfnamefont {R.}~\bibnamefont {Kosloff}},\
  }\href {\doibase 10.1103/PhysRevLett.108.070604} {\bibfield  {journal}
  {\bibinfo  {journal} {Phys. Rev. Lett.}\ }\textbf {\bibinfo {volume} {108}},\
  \bibinfo {pages} {070604} (\bibinfo {year} {2012}{\natexlab{b}})}\BibitemShut
  {NoStop}%
\bibitem [{\citenamefont {Uzdin}\ \emph {et~al.}(2015)\citenamefont {Uzdin},
  \citenamefont {Levy},\ and\ \citenamefont {Kosloff}}]{uzdin2015heatleak}%
  \BibitemOpen
  \bibfield  {author} {\bibinfo {author} {\bibfnamefont {R.}~\bibnamefont
  {Uzdin}}, \bibinfo {author} {\bibfnamefont {A.}~\bibnamefont {Levy}}, \ and\
  \bibinfo {author} {\bibfnamefont {R.}~\bibnamefont {Kosloff}},\ }\href
  {\doibase 10.1103/PhysRevX.5.031044} {\bibfield  {journal} {\bibinfo
  {journal} {Phys. Rev. X}\ }\textbf {\bibinfo {volume} {5}},\ \bibinfo {pages}
  {031044} (\bibinfo {year} {2015})}\BibitemShut {NoStop}%
\bibitem [{\citenamefont {Curzon}\ and\ \citenamefont
  {Ahlborn}(1975)}]{curzon1975efficiency}%
  \BibitemOpen
  \bibfield  {author} {\bibinfo {author} {\bibfnamefont {F.}~\bibnamefont
  {Curzon}}\ and\ \bibinfo {author} {\bibfnamefont {B.}~\bibnamefont
  {Ahlborn}},\ }\href {\doibase 10.1119/1.10023} {\bibfield  {journal}
  {\bibinfo  {journal} {Am. J. Phys.}\ }\textbf {\bibinfo {volume} {43}},\
  \bibinfo {pages} {22} (\bibinfo {year} {1975})}\BibitemShut {NoStop}%
\bibitem [{\citenamefont {Geva}\ and\ \citenamefont
  {Kosloff}(1994)}]{geva1994}%
  \BibitemOpen
  \bibfield  {author} {\bibinfo {author} {\bibfnamefont {E.}~\bibnamefont
  {Geva}}\ and\ \bibinfo {author} {\bibfnamefont {R.}~\bibnamefont {Kosloff}},\
  }\href {\doibase 10.1103/PhysRevE.49.3903} {\bibfield  {journal} {\bibinfo
  {journal} {Phys. Rev. E}\ }\textbf {\bibinfo {volume} {49}},\ \bibinfo
  {pages} {3903} (\bibinfo {year} {1994})}\BibitemShut {NoStop}%
\bibitem [{\citenamefont {Kosloff}\ and\ \citenamefont
  {Feldmann}(2002)}]{kosloff2002friction}%
  \BibitemOpen
  \bibfield  {author} {\bibinfo {author} {\bibfnamefont {R.}~\bibnamefont
  {Kosloff}}\ and\ \bibinfo {author} {\bibfnamefont {T.}~\bibnamefont
  {Feldmann}},\ }\href {\doibase 10.1103/PhysRevE.65.055102} {\bibfield
  {journal} {\bibinfo  {journal} {Phys. Rev. E}\ }\textbf {\bibinfo {volume}
  {65}},\ \bibinfo {pages} {055102} (\bibinfo {year} {2002})}\BibitemShut
  {NoStop}%
\bibitem [{\citenamefont {Feldmann}\ and\ \citenamefont
  {Kosloff}(2003)}]{feldmann2003friction}%
  \BibitemOpen
  \bibfield  {author} {\bibinfo {author} {\bibfnamefont {T.}~\bibnamefont
  {Feldmann}}\ and\ \bibinfo {author} {\bibfnamefont {R.}~\bibnamefont
  {Kosloff}},\ }\href {\doibase 10.1103/PhysRevE.68.016101} {\bibfield
  {journal} {\bibinfo  {journal} {Phys. Rev. E}\ }\textbf {\bibinfo {volume}
  {68}},\ \bibinfo {pages} {016101} (\bibinfo {year} {2003})}\BibitemShut
  {NoStop}%
\bibitem [{\citenamefont {Wang}\ \emph {et~al.}(2007)\citenamefont {Wang},
  \citenamefont {He},\ and\ \citenamefont {Xin}}]{wang2007friction}%
  \BibitemOpen
  \bibfield  {author} {\bibinfo {author} {\bibfnamefont {J.}~\bibnamefont
  {Wang}}, \bibinfo {author} {\bibfnamefont {J.}~\bibnamefont {He}}, \ and\
  \bibinfo {author} {\bibfnamefont {Y.}~\bibnamefont {Xin}},\ }\href {\doibase
  10.1088/0031-8949/75/2/018} {\bibfield  {journal} {\bibinfo  {journal} {Phys.
  Scr.}\ }\textbf {\bibinfo {volume} {75}},\ \bibinfo {pages} {227} (\bibinfo
  {year} {2007})}\BibitemShut {NoStop}%
\bibitem [{\citenamefont {\c{C}akmak}\ \emph {et~al.}(2016)\citenamefont
  {\c{C}akmak}, \citenamefont {Altintas},\ and\ \citenamefont {\"{O}zg\"{u}r
  E~M\"{u}stecapl\i{}o\u{g}lu}}]{cakmak2016}%
  \BibitemOpen
  \bibfield  {author} {\bibinfo {author} {\bibfnamefont {S.}~\bibnamefont
  {\c{C}akmak}}, \bibinfo {author} {\bibfnamefont {F.}~\bibnamefont
  {Altintas}}, \ and\ \bibinfo {author} {\bibnamefont {\"{O}zg\"{u}r
  E~M\"{u}stecapl\i{}o\u{g}lu}},\ }\href
  {http://stacks.iop.org/1402-4896/91/i=7/a=075101} {\bibfield  {journal}
  {\bibinfo  {journal} {Phys. Scr.}\ }\textbf {\bibinfo {volume} {91}},\
  \bibinfo {pages} {075101} (\bibinfo {year} {2016})}\BibitemShut {NoStop}%
\bibitem [{\citenamefont {Caldeira}\ and\ \citenamefont
  {Leggett}(1983)}]{caldeira1983path}%
  \BibitemOpen
  \bibfield  {author} {\bibinfo {author} {\bibfnamefont {A.}~\bibnamefont
  {Caldeira}}\ and\ \bibinfo {author} {\bibfnamefont {A.}~\bibnamefont
  {Leggett}},\ }\href {\doibase https://doi.org/10.1016/0378-4371(83)90013-4}
  {\bibfield  {journal} {\bibinfo  {journal} {Physica A}\ }\textbf {\bibinfo
  {volume} {121}},\ \bibinfo {pages} {587 } (\bibinfo {year}
  {1983})}\BibitemShut {NoStop}%
\bibitem [{\citenamefont {Di\'osi}(1993)}]{diosi1993calderia}%
  \BibitemOpen
  \bibfield  {author} {\bibinfo {author} {\bibfnamefont {L.}~\bibnamefont
  {Di\'osi}},\ }\href {\doibase 10.1016/0378-4371(93)90065-C} {\bibfield
  {journal} {\bibinfo  {journal} {Physica A}\ }\textbf {\bibinfo {volume}
  {199}},\ \bibinfo {pages} {517 } (\bibinfo {year} {1993})}\BibitemShut
  {NoStop}%
\bibitem [{\citenamefont {Vacchini}\ and\ \citenamefont
  {Hornberger}(2009)}]{vacchini2009}%
  \BibitemOpen
  \bibfield  {author} {\bibinfo {author} {\bibfnamefont {B.}~\bibnamefont
  {Vacchini}}\ and\ \bibinfo {author} {\bibfnamefont {K.}~\bibnamefont
  {Hornberger}},\ }\href {\doibase
  https://doi.org/10.1016/j.physrep.2009.06.001} {\bibfield  {journal}
  {\bibinfo  {journal} {Phys. Rep.}\ }\textbf {\bibinfo {volume} {478}},\
  \bibinfo {pages} {71 } (\bibinfo {year} {2009})}\BibitemShut {NoStop}%
\bibitem [{\citenamefont {Malabarba}\ \emph {et~al.}(2015)\citenamefont
  {Malabarba}, \citenamefont {Short},\ and\ \citenamefont
  {Kammerlander}}]{malabarba2015}%
  \BibitemOpen
  \bibfield  {author} {\bibinfo {author} {\bibfnamefont {A.~S.~L.}\
  \bibnamefont {Malabarba}}, \bibinfo {author} {\bibfnamefont {A.~J.}\
  \bibnamefont {Short}}, \ and\ \bibinfo {author} {\bibfnamefont
  {P.}~\bibnamefont {Kammerlander}},\ }\href
  {http://stacks.iop.org/1367-2630/17/i=4/a=045027} {\bibfield  {journal}
  {\bibinfo  {journal} {New J. Phys.}\ }\textbf {\bibinfo {volume} {17}},\
  \bibinfo {pages} {045027} (\bibinfo {year} {2015})}\BibitemShut {NoStop}%
\bibitem [{\citenamefont {Woods}\ \emph {et~al.}(2017)\citenamefont {Woods},
  \citenamefont {Silva},\ and\ \citenamefont {Oppenheim}}]{Woods2016}%
  \BibitemOpen
  \bibfield  {author} {\bibinfo {author} {\bibfnamefont {M.~P.}\ \bibnamefont
  {Woods}}, \bibinfo {author} {\bibfnamefont {R.}~\bibnamefont {Silva}}, \ and\
  \bibinfo {author} {\bibfnamefont {J.}~\bibnamefont {Oppenheim}},\ }\href
  {https://arxiv.org/abs/1607.04591} {\bibfield  {journal} {\bibinfo  {journal}
  {arXiv:1607.04591}\ } (\bibinfo {year} {2017})}\BibitemShut {NoStop}%
\bibitem [{\citenamefont {L{\"{o}}rch}\ \emph {et~al.}(2018)\citenamefont
  {L{\"{o}}rch}, \citenamefont {Bruder}, \citenamefont {Brunner},\ and\
  \citenamefont {Hofer}}]{Loerch2018}%
  \BibitemOpen
  \bibfield  {author} {\bibinfo {author} {\bibfnamefont {N.}~\bibnamefont
  {L{\"{o}}rch}}, \bibinfo {author} {\bibfnamefont {C.}~\bibnamefont {Bruder}},
  \bibinfo {author} {\bibfnamefont {N.}~\bibnamefont {Brunner}}, \ and\
  \bibinfo {author} {\bibfnamefont {P.~P.}\ \bibnamefont {Hofer}},\ }\href
  {http://arxiv.org/abs/1802.10572} {\bibfield  {journal} {\bibinfo  {journal}
  {arXiv:1802.10572}\ } (\bibinfo {year} {2018})}\BibitemShut {NoStop}%
\end{thebibliography}

%

\end{document}